\pdfoutput=1
\documentclass[10pt,a4paper]{article}
\usepackage[latin1]{inputenc}
\usepackage{amsmath}
\usepackage{amsfonts}
\usepackage{amssymb}
\usepackage{graphicx}
\usepackage{float}
\usepackage{setspace}
\usepackage{dsfont}		
\usepackage[all]{xy}	
\usepackage{epstopdf}
\usepackage[draft=false
,colorlinks=true
,linkcolor=blue
,citecolor=blue]{hyperref}
\usepackage{cleveref}
\usepackage{ifthen}  	
\usepackage{morefloats}

\usepackage{authblk}
\usepackage{natbib}
\setcitestyle{authoryear,open={(},close={)}}
\usepackage{amsthm}
\usepackage[margin=2.5cm]{geometry}

\floatstyle{ruled}
\newfloat{algorithm}{tbp}{loa}
\providecommand{\algorithmname}{Algorithm}
\floatname{algorithm}{\protect\algorithmname}

\newcommand{\Mod}[1]{\ (\text{mod}\ #1)}	

\makeatletter
\newcommand*{\textlabel}[1]{%
  \edef\@currentlabel{#1}
  \phantomsection
  \textbf{#1}
}
\makeatother

\newenvironment{prooflabel}{}{\itshape}{\rmfamily}
\renewenvironment{prooflabel}[1][\prooflabelname]
 {\renewcommand\proofname{\ifthenelse{\equal{#1}{}}{Proof.}{Proof of #1.}
 }\proof}
 {\endproof}
 
\newtheorem{theorem}{Theorem}[section]
\newtheorem{proposition}{Proposition}[section]
\newtheorem{lemma}{Lemma}[section]
\newtheorem{corollary}{Corollary}[section]
 
\theoremstyle{definition}
\newtheorem{definition}{Definition}[section]

\theoremstyle{remark}
\newtheorem{remark}{Remark}[section]

\begin{document}\sloppy

\title{Stability of Noisy Metropolis--Hastings}

\author{F. J. Medina-Aguayo}
\author{A. Lee}
\author{G. O. Roberts}
\affil{University of Warwick}

\maketitle

\begin{abstract}
Pseudo-marginal Markov chain Monte Carlo methods for sampling from 
intractable distributions have gained
recent interest and have been theoretically studied in considerable depth.
Their main appeal is that they are exact, in the sense that they target marginally
the correct invariant distribution. However, the pseudo-marginal
Markov chain can exhibit poor mixing and slow convergence towards its target. 
As an alternative, a subtly different Markov chain can be simulated, where
better mixing is possible but the exactness property is sacrificed. 
This is the noisy algorithm, initially conceptualised as Monte Carlo within
Metropolis (MCWM), which has also been studied but to a lesser extent. The present article provides a further characterisation of the noisy
algorithm, with a focus on fundamental stability properties like positive
recurrence and geometric ergodicity. Sufficient conditions for inheriting
geometric ergodicity from a standard Metropolis--Hastings chain are given, 
as well as convergence of the invariant distribution towards the 
true target distribution.
\medskip

\textbf{Keywords:} Markov chain Monte Carlo; Pseudo-marginal Monte Carlo; Monte Carlo within Metropolis; Intractable likelihoods; Geometric ergodicity.

\end{abstract}

\section{Introduction}

\subsection{Intractable target densities and the pseudo-marginal algorithm}

Suppose our aim is to simulate from an intractable probability
distribution $\pi$ for some random variable $X$, which takes values
in a measurable space $\left(\mathcal{X},\mathcal{B}(\mathcal{X})\right)$.
In addition, let $\pi$ have a density $\pi(x)$ with respect to some
reference measure $\mu(dx)$, e.g. the counting or the Lebesgue
measure. By intractable we mean that an analytical expression
for the density $\pi(x)$ is not available and so implementation
of a Markov chain Monte Carlo (MCMC) method targeting $\pi$ is not
straightforward. 

One possible solution to this problem is to target a different
distribution on the extended space $\left(\mathcal{X}\times\mathcal{W},\mathcal{B}(\mathcal{X})\times\mathcal{B}(\mathcal{W})\right)$,
which admits $\pi$ as marginal distribution. The pseudo-marginal 
algorithm (\citealt{Beaumont_2003}, \citealt{AndrieuNRoberts_2009})
falls into this category since it is a Metropolis--Hastings (MH) algorithm
targeting a distribution $\bar{\pi}_{N}$, associated to the random
vector $(X,W)$ defined on the product space $\left(\mathcal{X}\times\mathcal{W},\mathcal{B}(\mathcal{X})\times\mathcal{B}(\mathcal{W})\right)$
where $\mathcal{W}\subseteq\mathbb{R}^{+}_0:=[0,\infty)$. It is given
by
\begin{align}\label{eq:pi.bar}
\bar{\pi}_{N}(dx,dw) &:= \pi(dx)Q_{x,N}(dw)w,
\end{align}
where $\left\{ Q_{x,N}\right\} _{(x,N)\in\mathcal{X}\times\mathbb{N}^{+}}$
is a family of probability distributions on $\left(\mathcal{W},\mathcal{B}(\mathcal{W})\right)$
satisfying for each $(x,N)\in\mathcal{X}\times\mathbb{N}$ 
\begin{align}
\mathbb{E}_{Q_{x,N}}\left[W_{x,N}\right] & \equiv 1,\quad\text{for }W_{x,N}\sim Q_{x,N}(\cdot).\label{eq:weights.general}
\end{align}
Throughout this article, we restrict our attention to the case where for each $x\in\mathcal{X}$, $W_{x,N}$ is $Q_{x,N}$-a.s. strictly positive, for reasons that will become clear.

The random variables $\left\{ W_{x,N}\right\} _{x,N}$ are commonly
referred as the weights. Formalising this algorithm using \eqref{eq:pi.bar} and \eqref{eq:weights.general} was introduced by \citet{AndrieuNVihola_2015}, and ``exactness'' follows immediately: $\bar{\pi}$ admits $\pi$ as a marginal. Given a proposal kernel $q:\mathcal{X}\times\mathcal{B}(\mathcal{X})\rightarrow[0,1]$, the respective proposal of the pseudo-marginal is given by 
\begin{align*}
\bar{q}_{N}(x,w;dy,du) &:= q(x,dy)Q_{y,N}(du),
\end{align*}
and, consequently, the acceptance probability can be expressed as
\begin{align}\label{eq:alpha.bar}
\bar{\alpha}_{N}(x,w;y,u) &:= \min\left\{ 1,\frac{\pi(dy)uq(y,dx)}{\pi(dx)wq(x,dy)}\right\} .
\end{align}
The pseudo-marginal algorithm defines a time-homogeneous
Markov chain, with transition kernel $\bar{P}_{N}$ on the measurable
space $\left(\mathcal{X}\times\mathcal{W},\mathcal{B}(\mathcal{X})\times\mathcal{B}(\mathcal{W})\right)$.
A single draw from $\bar{P}_{N}(x,w;\cdot,\cdot)$
is presented in \Cref{alg:PM}.

\begin{algorithm}[ht]
\begin{singlespace}
\caption{\label{alg:PM}Simulating from $\bar{P}_{N}(x,w;\cdot,\cdot)$}
\end{singlespace}
\begin{enumerate}
\item Sample $Y\sim q(x,\cdot)$.
\item Draw $U\sim Q_{Y,N}(\cdot)$.
\item With probability $\bar{\alpha}_{N}(x,w;Y,U)$ defined in \eqref{eq:alpha.bar}:\\
\hphantom{xx} return $(Y,U)$,\\
otherwise:\\
\hphantom{xx} return $(x,w)$.
\end{enumerate}
\end{algorithm}

Due to its exactness and straightforward implementation
in many settings, the pseudo-marginal has gained recent interest and
has been theoretically studied in some depth, see e.g. \citet{AndrieuNRoberts_2009}, \citet{AndrieuNVihola_2015}, \citet{AndrieuNVihola_2014}, \citet{SherlockETal_2015},
\citet{Girolami_2013} and \citet{MaireETal_2014}. These studies typically compare the pseudo-marginal Markov chain with a ``marginal'' Markov chain, arising in the case where all the weights are almost surely equal to $1$, and \eqref{eq:alpha.bar} is then the standard Metropolis--Hastings acceptance probability associated with the target density $\pi$ and the proposal $q$.

\subsection{Examples of pseudo-marginal algorithms}

A common source of intractability for $\pi$ occurs when a latent variable $Z$ on $(Z,\mathcal{B}(Z))$ is used to model observed data, as in hidden Markov models (HMMs) or mixture models. Although the density $\pi(x)$ cannot be computed, it can be approximated via importance sampling, using an appropriate auxiliary distribution, say $\nu_{x}$.
Here, appropriate means $\pi_{x}\ll\nu_{x}$, where $\pi_{x}$ denotes
the conditional distribution of $Z$ given $X=x$. Therefore, for this setting,
the weights are given by 
\begin{align*}
W_{x,N} &= \frac{1}{N}\sum_{k=1}^{N}\frac{\pi_{x}\left(Z_{x}^{(k)}\right)}{\nu_{x}\left(Z_{x}^{(k)}\right)}, \quad \text{where }\left\{ Z_{x}^{(k)}\right\} _{k\in\{1,\dots,N\}}\overset{i.i.d.}{\sim}\nu_{x}(\cdot),
\end{align*}
which motivates the following generic form when using averages of
unbiased estimators 
\begin{align}
\begin{split}W_{x,N} &= \frac{1}{N}\sum_{k=1}^{N}W_{x}^{(k)}, \quad \text{where }\left\{ W_{x}^{(k)}\right\} _{k}\overset{i.i.d.}{\sim}Q_{x}(\cdot),\mathbb{E}_{Q_{x}}\left[W_{x}^{(k)}\right]\equiv1.
\end{split}
\label{eq:wghts.arithm.avg}
\end{align}
It is clear that \eqref{eq:wghts.arithm.avg} describes only a special case
of \eqref{eq:weights.general}. Nevertheless, we will pay special attention to the former throughout the article. For similar settings to \eqref{eq:wghts.arithm.avg} see \citet{AndrieuNRoberts_2009}. 

\begin{figure}[ht]
\centering $\xymatrix{ & Y_{1} & & Y_{t} & & Y_{T} \\ X_0 \ar[r]^{f_\theta} & X_1 \ar[u]^{g_\theta} \ar[r]^{f_\theta} & \dots \ar[r]^{f_\theta} & X_{t} \ar[u]^{g_\theta} \ar[r]^{f_\theta} & \dots \ar[r]^{f_\theta} & X_{T} \ar[u]^{g_\theta} }$
\caption{Hidden Markov Model.}
\label{fig:HMM}
\end{figure}
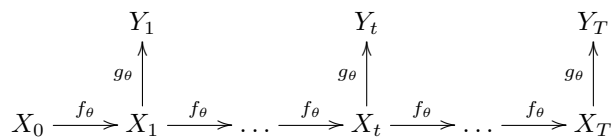

Since \eqref{eq:weights.general} is more general, it allows $W_{x,N}$ to be any random variable with expectation 1. Sequential Monte Carlo (SMC) methods involve the simulation of a system of some number of particles, and provide unbiased estimates of likelihoods associated with HMMs (see \citealt[][Proposition~7.4.1]{MoralFeynman_2004} or \citealt{PittETal_2012}) irrespective of the size of the particle system.  Consider the model given by \Cref{fig:HMM}. 
The random variables $\left\{ X_{t}\right\} _{t=0}^{T}$ form a time-homogeneous Markov chain with transition $f_{\theta}(\cdot|x_{t-1})$ that depends on a set of parameters $\theta$. The observed random
variables $\left\{ Y_{t}\right\} _{t=1}^{T}$ are conditionally independent
given the unobserved $\left\{ X_{t}\right\} _{t=1}^{T}$ and are distributed according to $g_{\theta}(\cdot|x_{t})$, which also may depend on $\theta$. The likelihood function for $\theta$ is given by
\begin{align*}
l(\theta;y_1,\dots,y_T):= \mathbb{E}_{f_\theta} \left [\prod_{t=1}^T g_\theta(y_{t}|X_{t}) \right],
\end{align*}
where $\mathbb{E}_\theta$ denotes expectation w.r.t. the $\theta$-dependent law of $\{X_t\}_{t=1}^T$, and we assume for simplicity that the initial value $X_0=x_0$ is known.
If we denote by $\hat{l}_N(\theta;y_1,\dots,y_T)$ the unbiased SMC estimator of $l(\theta;y_1,\dots,y_T)$ based on $N$ particles, we can then define
\begin{align*}
W_{\theta,N}:= \frac{\hat{l}_N(\theta;y_1,\dots,y_T)}{l(\theta;y_1,\dots,y_T)},
\end{align*}
and \eqref{eq:weights.general} is satisfied but \eqref{eq:wghts.arithm.avg} is not.
The resulting pseudo-marginal algorithm is developed and discussed in detail in \citet{AndrieuETal_2010}, where it and related algorithms are referred to as particle MCMC methods.

\subsection{The noisy algorithm}

Although the pseudo-marginal has the desirable property of exactness,
it can suffer from ``sticky'' behaviour, exhibiting poor mixing and
slow convergence towards the target distribution (\citealt{AndrieuNRoberts_2009} and 
\citealt{LeeNLatuszynski_2014}).
The cause for this is well-known to be related with the value of the
ratio between $W_{y,N}$ and $W_{x,N}$ at a particular iteration.
Heuristically, when the value of the current weight ($w$ in \eqref{eq:alpha.bar})
is large, proposed moves can have a low probability of acceptance. As a consequence, the resulting chain can get ``stuck'' and may not move after a considerable number of iterations.

In order to overcome this issue, a
subtly different algorithm is performed in some practical problems \citep[see, e.g.,][]{McKinleyETal_2014}.
The basic idea is to refresh, independently from the past, the value
of the current weight at every iteration. The
ratio of the weights between $W_{y,N}$ and $W_{x,N}$ still plays
an important role in this alternative algorithm, but here refreshing
$W_{x,N}$ at every iteration can improve mixing and the rate of convergence.

This alternative algorithm is commonly known as Monte Carlo within
Metropolis (MCWM), as in \citet{ONeillETal_2000}, \citet{Beaumont_2003} or \citet{AndrieuNRoberts_2009}, since typically the weights are Monte Carlo estimates as in \eqref{eq:wghts.arithm.avg}. From this point onwards it will be referred as the noisy MH algorithm
or simply the noisy algorithm to emphasize that our main assumption is \eqref{eq:weights.general}. Due to independence from previous iterations
while sampling $W_{x,N}$ and $W_{y,N}$, the noisy algorithm also
defines a time-homogeneous Markov chain with transition kernel
$\tilde{P}_{N}$, but on the measurable space $(\mathcal{X},\mathcal{B}(\mathcal{X}))$. A single draw from $\tilde{P}_{N}(x,\cdot)$ is presented in \Cref{alg:nMH}, and it is clear that we restrict our attention to strictly positive weights because the algorithm is not well-defined when both $W_{y,N}$ and $W_{x,N}$ are equal to $0$. 

\begin{algorithm}[ht]
\caption{\label{alg:nMH}Simulating from $\tilde{P}_{N}(x,\cdot)$}
\begin{enumerate}
\item Sample $Y\sim q(x,\cdot)$.
\item Draw $W\sim Q_{x,N}(\cdot)$ and $U\sim Q_{Y,N}(\cdot)$, independently.
\item With probability $\bar{\alpha}_{N}(x,W;Y,U)$ defined in \eqref{eq:alpha.bar}: \\
\hphantom{xx} return $Y$,\\
otherwise:\\
\hphantom{xx} return $x$.
\end{enumerate}
\end{algorithm}

Even though these algorithms differ only slightly, the related chains
have very different properties. In \Cref{alg:nMH},
the value $w$ is generated at every iteration whereas in \Cref{alg:PM}, 
it is treated as an input. As a consequence, \Cref{alg:PM} produces a chain on $\left(\mathcal{X}\times\mathcal{W},\mathcal{B}(\mathcal{X})\times\mathcal{B}(\mathcal{W})\right)$
contrasting with a chain from \Cref{alg:nMH} taking
values on $\left(\mathcal{X},\mathcal{B}(\mathcal{X})\right)$. However,
the noisy chain is not invariant under $\pi$ and it is not reversible
in general. Moreover, it may not even have an invariant distribution
as shown by some examples in \Cref{sec:Examples}.

From \citet{ONeillETal_2000} and \citet{FerndzNRubio_2007}, it is evident that
the implementation of the noisy algorithm goes back even before the
appearance of the pseudo-marginal, the latter initially conceptualised as Grouped
Independence Metropolis--Hastings (GIMH) in \citet{Beaumont_2003}.
Theoretical properties, however, of the noisy algorithm have mainly been
studied in tandem with the pseudo-marginal by \citet{Beaumont_2003}, \citet{AndrieuNRoberts_2009} and more recently by \citet{EverittETal_2014}.

\subsection{Objectives of the article}

\begin{figure}
\centering
\includegraphics[width=0.96\linewidth]{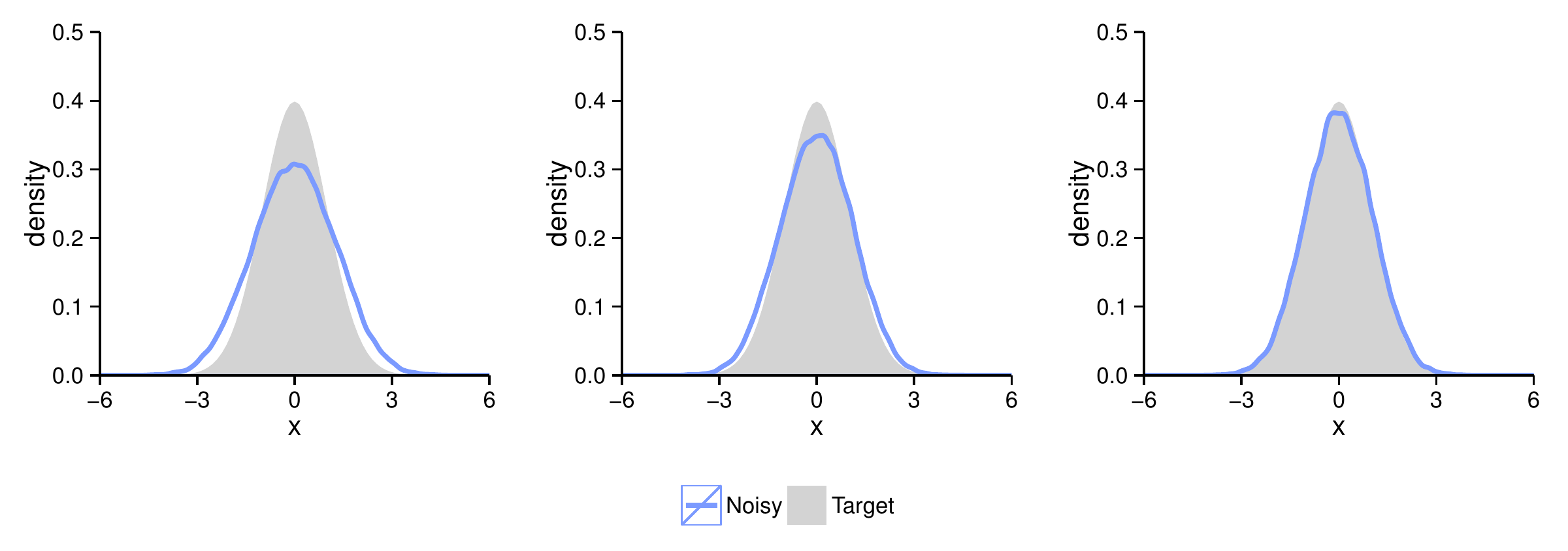}
\caption{Estimated densities using the noisy
chain with $100,000$ iterations for $N=10$ (left), $N=100$ (central) and $N=1,000$ (right).}
\label{fig:logN.densities}
\end{figure}

The objectives of this article can be illustrated using a simple example. Let $\mathcal{N}(\cdot | \mu, \sigma^2)$ denote a univariate Gaussian distribution with mean $\mu$ and variance $\sigma^2$ and $\pi(\cdot)=\mathcal{N}(\cdot | 0, 1)$ be a standard normal distribution. Let the weights $W_{x,N}$ be as in \eqref{eq:wghts.arithm.avg} with
\begin{align*}
Q_x(\cdot)=\log \mathcal{N}\left( \cdot \Big|-\frac{1}{2}\sigma^2, \sigma^2 \right) \quad \text{and} \quad \sigma^2:=5,
\end{align*}
where $\log\mathcal{N}(\cdot | \mu, \sigma^2)$ denotes a log-normal distribution of parameters $\mu$ and $\sigma^2$.
In addition, let the proposal $q$ be random walk given by $q(x,\cdot)=\mathcal{N}\left( \cdot |x, 4 \right)$.
For this example, \Cref{fig:logN.densities} shows the estimated densities using the noisy chain for different values of $N$. It appears that the noisy chain has an invariant distribution, and as $N$ increases it seems to approach the desired target $\pi$. Our objectives here are to answer the following types of questions about the noisy algorithm in general:
\begin{enumerate}
\item{Does an invariant distribution exist, at least for $N$ large enough?}
\item{Does the noisy Markov chain behave like the marginal chain for sufficiently large $N$?}
\item{Does the invariant distribution, if it exists, converge to $\pi$ as $N$ increases?}
\end{enumerate}
We will see that the answer to the first two questions is negative in general. However, all three questions can be answered positively when the marginal chain is geometrically ergodic and the distributions of the weights satisfy additional assumptions.

\subsection{Marginal chains and geometric ergodicity}

In order to formalise our analysis, let $P$ denote the Markov transition
kernel of a standard MH chain on $\left(\mathcal{X},\mathcal{B}(\mathcal{X})\right)$,
$ $targeting $\pi$ with proposal $q$. We will refer to this chain
and this algorithm using the term marginal (as in \citealt{AndrieuNRoberts_2009} and \citealt{AndrieuNVihola_2015}),
which can be seen as an idealised version for which the noisy chain
and corresponding algorithm are simple approximations. Therefore 
\begin{align*}
P(x,dy) &:= \alpha(x,y)q(x,dy)+\delta_{x}(dy)\rho(x),
\end{align*}
where $\alpha$ is the MH acceptance probability and $\rho$ is the
rejection probability, given by
\begin{align}
\alpha(x,y):=\min\left\{ 1,\frac{\pi(dy)q(y,dx)}{\pi(dx)q(x,dy)}\right\} \quad\text{and} \quad \rho(x):=1-\int_{\mathcal{X}}\alpha(x,y)q(x,dy).
\label{eq:alpha_rho}
\end{align}
Similarly, for the transition kernel $\tilde{P}_{N}$ of the noisy
chain, moves are proposed according to $q$ but are accepted using
$\bar{\alpha}_{N}$ (as in \eqref{eq:alpha.bar}) instead of $\alpha$,
once values for $W_{x,N}$ and $W_{y,N}$ are sampled. In order to
distinguish the acceptance probabilities between the noisy and the
pseudo-marginal processes, despite being the same after sampling values
for the weights, define 
\begin{align}
\tilde{\alpha}_{N}(x,y) &:= \mathbb{E}_{Q_{x,N}\otimes Q_{y,N}}\bar{\alpha}_{N}(x,W_{x,N};y,W_{y,N}).\label{eq:alpha_tilde}
\end{align}
Here $\tilde{\alpha}_{N}$ is the expectation of a randomised acceptance
probability, which permits defining the transition kernel of the noisy
chain by 
\begin{align*}
\tilde{P}_{N}(x,dy) &:= \tilde{\alpha}_{N}(x,y)q(x,dy)+\delta_{x}(dy)\tilde{\rho}_{N}(x),
\end{align*}
where $\tilde{\rho}_{N}$ is the noisy rejection probability given
by 
\begin{align}
\tilde{\rho}_{N}(x) &:= 1-\int_{\mathcal{X}}\tilde{\alpha}_{N}(x,y)q(x,dy).\label{eq:rho_tilde}
\end{align}
The noisy kernel $\tilde{P}_{N}$ is just a perturbed version of $P$, involving a ratio of weights in the noisy acceptance probability $\tilde{\alpha}_{N}$. In addition, when such weights are identically one, i.e. $Q_{x,N}(\{1\})=1$, the noisy chain reduces to the marginal chain, whereas the pseudo-marginal becomes the marginal chain with an extra component always equal to 1.

So far, the terms slow convergence and ``sticky'' behaviour have been
used in a relative vague sense. A powerful characterisation of the behaviour of a Markov chain is provided by geometric ergodicity, defined below. Geometrically ergodic Markov chains have a limiting invariant probability distribution, which they converge towards geometrically fast in total variation \citep{MeynNTweedie_2009}. For any Markov kernel $K:\mathcal{X}\times\mathcal{B}(\mathcal{X})\rightarrow[0,1]$,
let $K^{n}$ be the $n$-step transition kernel, which is given by
\begin{align*}
K^{n} & (x,\cdot):=\int_{\mathcal{X}}K^{n-1}(x,dz)K(z,\cdot),\quad\text{for }n\geq2.
\end{align*}

\begin{definition}[Geometric ergodicity]\label{defn:geom.ergo}
A $\varphi$-irreducible and aperiodic Markov chain $\mathbf{\Phi}:=(\Phi_i)_{i\geq 0}$ on a
measurable space $\left(\mathcal{X},\mathcal{B}(\mathcal{X})\right)$,
with transition kernel $P$ and invariant distribution $\pi$, is geometrically
ergodic if there exists a finite function $V\geq1$ and constants
$\tau<1$, $R<\infty$ such that
\begin{align}\label{eq:geom.ergo.condn}
\|P^{n}(x,\cdot)-\pi(\cdot)\|_{TV} &\leq RV(x)\tau^{n},\quad \text{for }x\in\mathcal{X}.
\end{align}
Here, $\|\cdot\|_{TV}$ denotes the total variation norm given by 
\begin{align*}
\|\mu\|_{TV} &= \frac{1}{2} \sup_{|g|\leq1}\Big|\int\mu(dy)g(y)\Big|=\sup_{A\in\mathcal{B}(\mathcal{X})}\mu(A),
\end{align*}
where $\mu$ is any signed measure.
\end{definition}
Geometric ergodicity does not necessarily provide fast convergence in an absolute sense. For instance, consider cases where $\tau$, or $R$, from \Cref{defn:geom.ergo} are extremely close to one, or very large respectively. Then the decay of the total variation distance, though geometric, is not particularly
fast (see \citealt{RobertsNRosenthal_2004} for some examples).

Nevertheless, geometric ergodicity is a useful tool when analysing non-reversible
Markov chains as will become apparent in the noisy chain case. Moreover, in practice one is often interested in estimating $\mathbb{E}_{\pi}\left[f(X)\right]$ for some function $f:\mathcal{X}\rightarrow\mathbb{R}$, which is done by using ergodic averages of the form 
\begin{align*}
e_{n,m}(f) &:= \frac{1}{n}\sum_{i=m+1}^{m+n}f\left(\Phi_{i}\right), \quad \text{for }m,n\geq0.
\end{align*}
In this case, geometric ergodicity is a desirable property since it
can guarantee the existence of a central limit theorem (CLT) for $e_{n,m}(f)$,
see \citet{ChanNGeyer_1994}, \citet{RobertsNRosenthal_1997} and \citet{RobertsNRosenthal_2004} for a more general review. Also, its importance is related with the construction of consistent estimators of the corresponding asymptotic
variance in the CLT, as in \citet{FlegalNJones_2010}.

As noted in \citet{AndrieuNRoberts_2009}, if the weights $\left\{ W_{x,N}\right\} _{x,N}$
are not essentially bounded then the pseudo-marginal chain cannot
be geometrically ergodic; in such cases the ``stickiness'' may be
more evident. In addition, under mild assumptions (in particular, that $\bar{P}_N$ has a left spectral gap), from \citet[][Proposition 10]{AndrieuNVihola_2015} and \citet{LeeNLatuszynski_2014},
a sufficient but not necessary condition ensuring the pseudo-marginal
inherits geometric ergodicity from the marginal, is that the weights are uniformly
bounded. This certainly imposes a tight restriction in many practical
problems.

The analyses in \citet{AndrieuNRoberts_2009} and \citet{EverittETal_2014} mainly study the noisy algorithm in the case where the marginal Markov chain is uniformly ergodic, i.e. when it satisfies \eqref{eq:geom.ergo.condn} with $\sup_{x\in\mathcal{X}} V(x)<\infty$.
However, there are many Metropolis--Hastings Markov chains for statistical estimation that cannot be uniformly ergodic, e.g. random walk Metropolis chains when $\pi$ is not compactly supported. Our focus is therefore on inheritance of geometric ergodicity by the noisy chain, complementing existing results for the pseudo-marginal chain.

\subsection{Outline of the paper}

In \Cref{sec:Examples}, some simple examples are presented
for which the noisy chain is positive recurrent, so it has
an invariant probability distribution. This is perhaps the weakest stability property that one would expect a Monte Carlo Markov chain to have. However, other fairly surprising examples are presented for which the noisy Markov chain is transient even though the marginal and pseudo-marginal chains
are geometrically ergodic. \Cref{sec:Inheritance} is dedicated
to inheritance of geometric ergodicity from the marginal chain,
where two different sets of sufficient conditions are given and are further analysed
in the context of arithmetic averages given by \eqref{eq:wghts.arithm.avg}.
Once geometric ergodicity is attained, it guarantees the existence
of an invariant distribution $\tilde{\pi}_{N}$ for the noisy chain.
Under the same sets of conditions, we show in \Cref{sec:Invariants} that $\tilde{\pi}_{N}$ and $\pi$ can be made arbitrarily close in total variation as $N$ increases. 
Moreover, explicit rates of convergence are possible to obtain in principle, when the weights arise from an arithmetic average setting as in \eqref{eq:wghts.arithm.avg}.

\section{Motivating examples\label{sec:Examples}}

\subsection{Homogeneous weights with a random walk proposal\label{sub:Hom.weights.rwm}}

Assume a log-concave target distribution $\pi$ on the positive integers,
whose density with respect to the counting measure is given by
\begin{align*}
\pi(m) &\propto \exp\left\{ -h(m)\right\} \mathds{1}_{m\in\mathbb{N}^{+}},
\end{align*}
where $h:\mathbb{N}^{+}\rightarrow\mathbb{R}$ is a convex function.
In addition, let the proposal distribution be a symmetric random walk
on the integers, i.e.
\begin{align}
q(m,\{m+1\})=  \frac{1}{2}=q(m,\{m-1\}),\quad\text{for }m\in\mathbb{Z}.\label{eq:sym.proposal}
\end{align}
From \citet{MengersenNTweedie_1996}, it can be seen that the marginal
chain is geometrically ergodic.

Now, assume the distribution of the weights $\left\{ W_{m,N}\right\} _{m,N}$
is homogeneous with respect to the state space, meaning
\begin{align}
W_{m,N}=W_{N} & \sim Q_{N}(\cdot),\quad\text{for all }m\in\mathbb{N}^{+}.\label{eq:ex.hom.weights}
\end{align}
In addition, assume $W_{N}>0$ $Q_{N}$-a.s., then for $m\geq2$
\begin{align*}
\tilde{P}_{N}(m,\left\{ m-1\right\} ) &= \frac{1}{2}\mathbb{E}_{Q_{N}\otimes Q_{N}}\left[\min\left\{ 1,\frac{\exp\{h(m)\}}{\exp\{h(m-1)\}}\times\frac{W_{N}^{(1)}}{W_{N}^{(2)}}\right\} \right] \quad \text{and}\\
\tilde{P}_{N}(m,\left\{ m+1\right\} ) &= \frac{1}{2}\mathbb{E}_{Q_{N}\otimes Q_{N}}\left[\min\left\{ 1,\frac{\exp\{h(m)\}}{\exp\{h(m+1)\}}\times\frac{W_{N}^{(1)}}{W_{N}^{(2)}}\right\} \right], \quad \text{where }\left\{ W_{N}^{(k)}\right\} _{k\in\{1,2\}}\overset{i.i.d.}{\sim}Q_N(\cdot).
\end{align*}
For this particular class of weights and using the fact that $h$
is convex, the noisy chain is geometrically ergodic, implying the
existence of an invariant probability distribution.
\begin{proposition}
\label{prop:ex.logconc.hom}Consider a log-concave target density
on the positive integers and a proposal density as in \eqref{eq:sym.proposal}.
In addition, let the distribution of the weights be homogeneous as
in \eqref{eq:ex.hom.weights}. Then, the chain generated by the noisy kernel $\tilde{P}_{N}$ is geometrically ergodic.
\end{proposition}
It is worth noting that the distribution of the weights, though homogeneous
with respect to the state space, can be taken arbitrarily, as long
as the weights are positive. Homogeneity ensures that the distribution
of the ratio of such weights is not concentrated near 0,
due to its symmetry around one, i.e. for $z>0$
\begin{align*}
\mathbb{P}_{Q_{N}\otimes Q_{N}}\left[\frac{W_{N}^{(1)}}{W_{N}^{(2)}}\leq z\right] &= \mathbb{P}_{Q_{N}\otimes Q_{N}}\left[\frac{W_{N}^{(1)}}{W_{N}^{(2)}}\geq\frac{1}{z}\right], \quad \text{where } \left\{ W_{N}^{(k)}\right\} _{k\in\{1,2\}}\overset{i.i.d.}{\sim}Q_{N}(\cdot).
\end{align*}
In contrast, when the support of the distribution $Q_{N}$ is unbounded, the corresponding
pseudo-marginal chain cannot be geometrically ergodic.

\subsection{Particle MCMC\label{sub:PMCMC}}

More complex examples arise when using particle MCMC methods,
for which noisy versions can also be performed. They may prove
to be useful in some inference problems. Consider again the hidden Markov model given by \Cref{fig:HMM}. As before, set $X_{0}=x_{0}$ and let
\begin{align*}
 & \theta=\left\{ x_{0},a,\sigma_{X}^{2},\sigma_{Y}^{2}\right\} ,\\
 & f_{\theta}\left(\cdot|X_{t-1}\right)=\mathcal{N}\left(\cdot|aX_{t-1},\sigma_{X}^{2}\right)\quad\text{and}\\
 & g_{\theta}\left(\cdot|X_{t}\right)=\mathcal{N}\left(\cdot|X_{t},\sigma_{Y}^{2}\right).
\end{align*}
Therefore,
once a prior distribution for $\theta$ is specified, $p(\cdot)$ say, the aim is to
conduct Bayesian inference on the posterior distribution
\begin{align*}
\pi(\theta|y_1,\dots,y_T)\propto p(\theta) l(\theta;y_1,\dots,y_T).
\end{align*}

\begin{figure}
\centering
\includegraphics[width=0.48\linewidth]{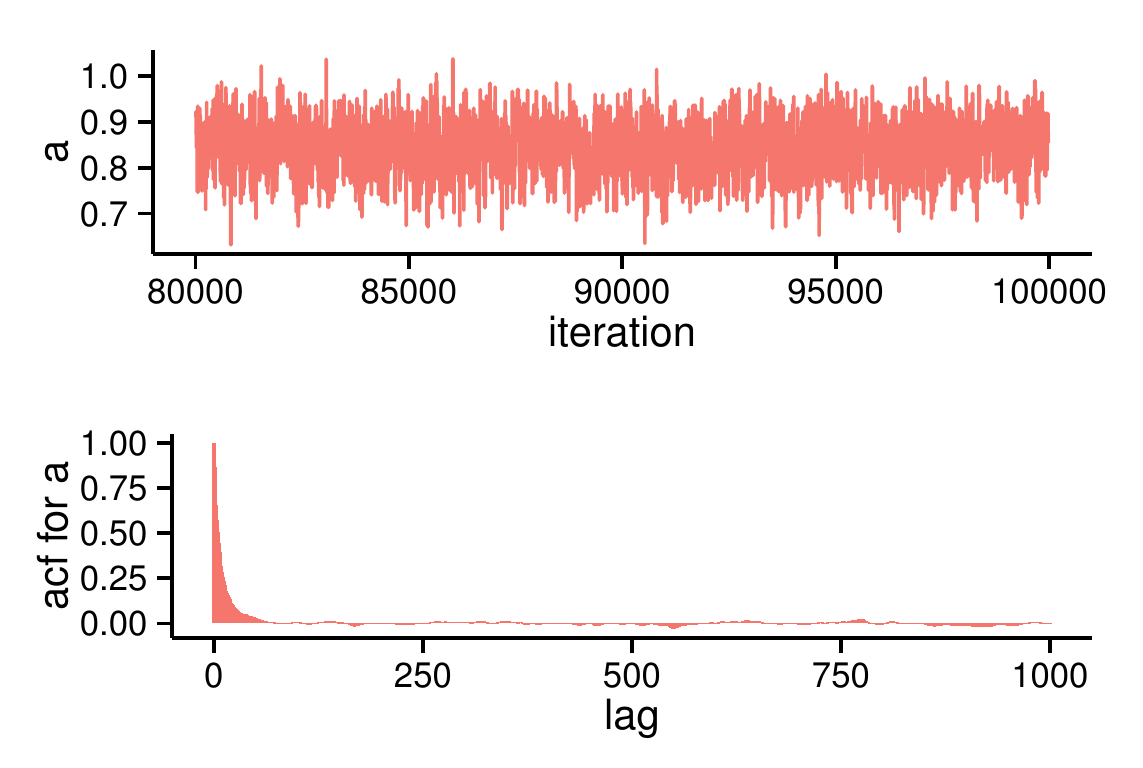}
\caption{Last $20,000$ iterations of the marginal algorithm for the autoregressive
parameter $a$ (top). Estimated autocorrelation function of the corresponding
marginal chain (bottom). The mean acceptance probability was $0.256$.}
\label{fig:PMMH.marginal}
\end{figure}

\begin{figure}
\centering
\includegraphics[width=0.96\linewidth]{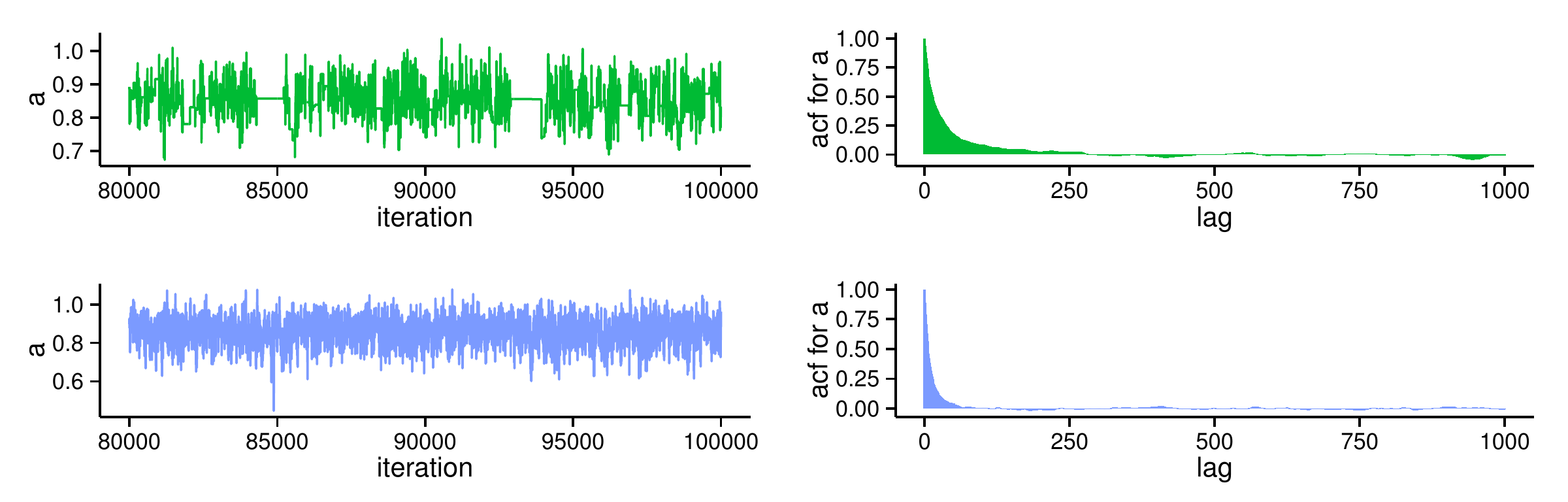}
\caption{Last $20,000$ iterations of the pseudo-marginal (top left) and noisy (bottom left)
algorithms, for the autoregressive parameter $a$ when $N=250$. Estimated
autocorrelation functions of the corresponding pseudo-marginal (top right)
and noisy (bottom right) chains. The mean acceptance probabilities were $0.104$ for the pseudo-marginal
and $0.283$ for the noisy chain.}
\label{fig:PMMH.PM_Noisy}
\end{figure}

In this particular setting, the posterior distribution is tractable. This will allows us to compare the results obtained from the exact and noisy versions, both relying on the SMC estimator $\hat{l}_N(\theta;y_1,\dots,y_T)$ of the likelihood. Using a uniform prior for the parameters and a random walk proposal,
\Cref{fig:PMMH.marginal} shows the run and autocorrelation
function (acf) for the autoregressive parameter $a$ of the marginal
chain.
Similarly, \Cref{fig:PMMH.PM_Noisy} shows the corresponding
run and acf for both the pseudo-marginal and the noisy chain when $N=250$. It is noticeable how the pseudo-marginal
gets ``stuck'', resulting in a lower acceptance than the marginal
and noisy chains. In addition, the acf of the noisy chain seems to
decay faster than that of the pseudo-marginal chain. 

\begin{figure}
\centering
\resizebox{0.8\hsize}{!}{\includegraphics*{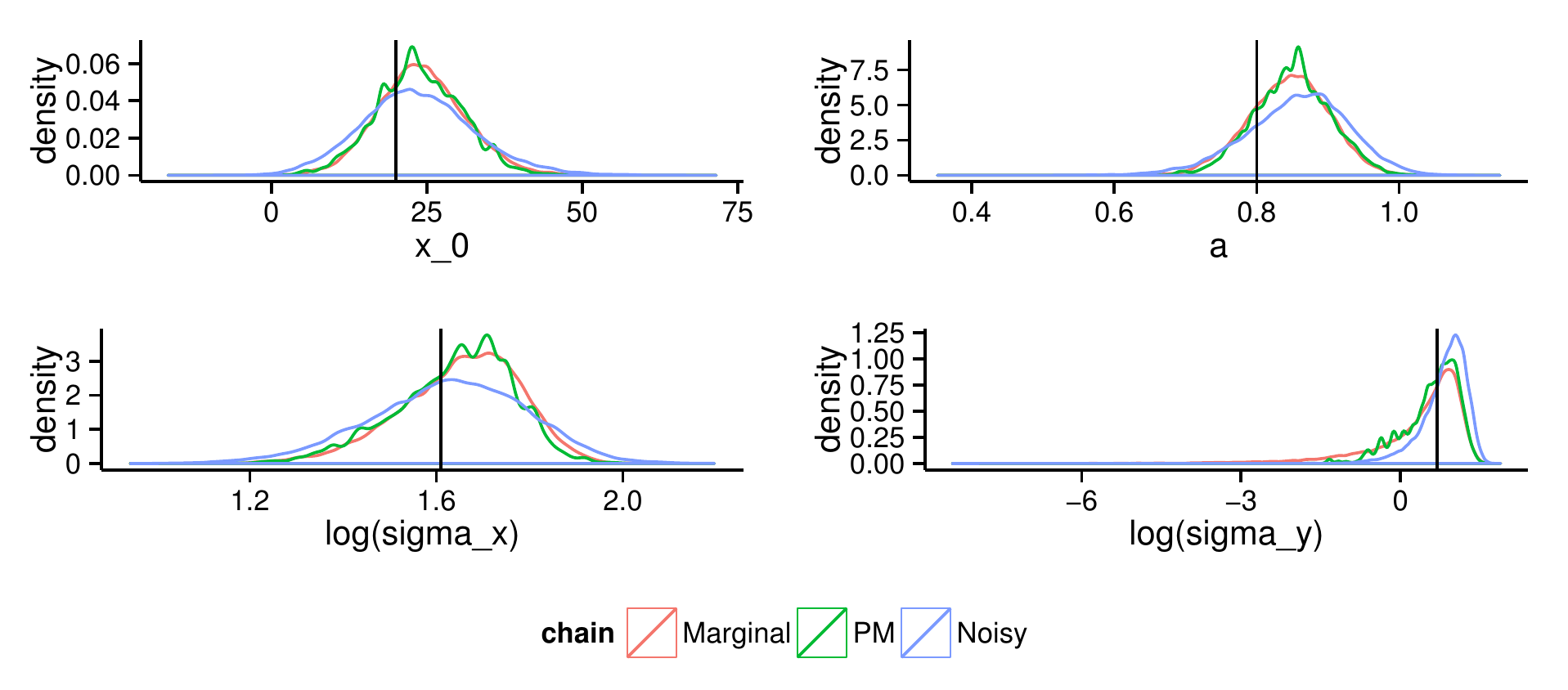}}
\caption{Estimated densities using the marginal, pseudo-marginal and noisy
chains for the 4 parameters when $N=250$. Vertical lines indicate the real values.}
\label{fig:PMMH.dens.250}
\end{figure}

\begin{figure}
\centering
\resizebox{0.8\hsize}{!}{\includegraphics*{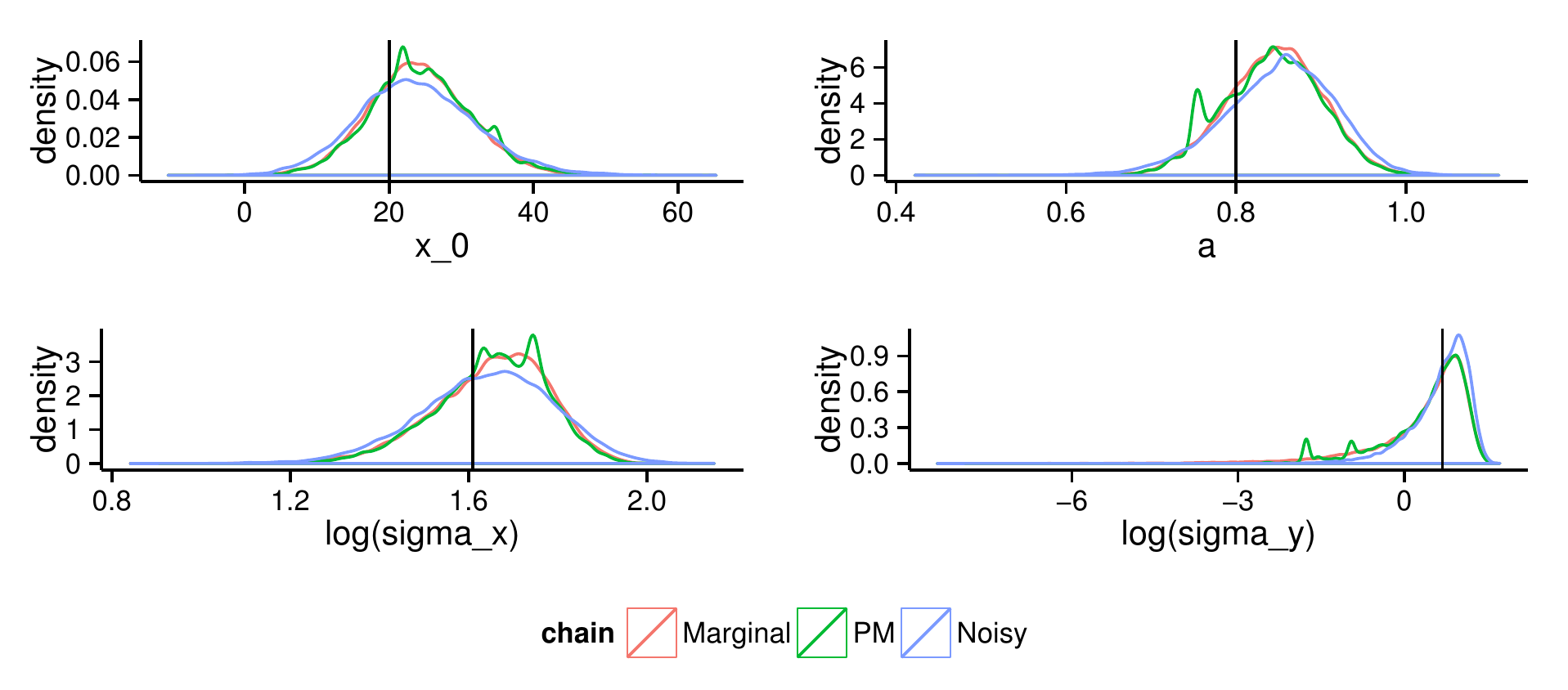}}
\caption{Estimated densities using the marginal, pseudo-marginal and noisy
chains for the 4 parameters, when $N=750$. Vertical lines indicate the real values.}
\label{fig:PMMH.dens.750}
\end{figure}

Finally, \Cref{fig:PMMH.dens.250} and \Cref{fig:PMMH.dens.750} show the estimated 
posterior densities for the parameters when $N=250$ and $N=750$, respectively. There, the trade-off 
between the pseudo-marginal and the noisy algorithm is noticeable. For lower values of $N$, the
pseudo-marginal will require more iterations due to the slow mixing,
whereas the noisy converges faster towards an unknown noisy invariant distribution.
By increasing $N$, the mixing in the pseudo-marginal improves and
the noisy invariant approaches the true posterior.

\subsection{Transient noisy chain with homogeneous weights\label{sub:Trans.hom.weights}}

In contrast with example in \Cref{sub:Hom.weights.rwm},
this one shows that the noisy algorithm can produce a transient chain
even in simple settings. Let $\pi$ be a geometric distribution on
the positive integers, whose density with respect to the counting
measure is given by
\begin{align}
\pi(m) &= \left(\frac{1}{2}\right)^{m}\mathds{1}_{\left\lbrace m\in\mathbb{N}^{+}\right\rbrace }.\label{eq:ex.geom.density}
\end{align}
In addition, assume the proposal distribution is a simple random walk
on the integers, i.e.
\begin{align}
q(m,\{m+1\}) = \theta=1-q(m,\{m-1\}),\quad\text{for }m\in\mathbb{Z}.\label{eq:ex.non.sym.prop}
\end{align}
where $\theta\in(0,1)$. Under these assumptions, the marginal chain
is geometrically ergodic, see \Cref{prop:non.hom.rw} in \Cref{sec:Proofs}.

Consider $N=1$ and as in \Cref{sub:Hom.weights.rwm}, let the
distribution of weights be homogeneous and given by
\begin{align}
W &= (b-\varepsilon)Ber(s)+\varepsilon,\quad\text{for }b>1\text{ and }\varepsilon\in(0,1),\label{eq:ex.Blli.weights}
\end{align}
where $Ber(s)$ denotes a Bernoulli random variable of parameter $s\in(0,1).$
There exists a relationship between
$s$, $b$ and $\varepsilon$ that guarantees the expectation of
the weights is identically one. The following proposition, proven in \Cref{sec:Proofs} by taking $\theta > 1/2$, shows that the resulting
noisy chain can be transient for certain values of $b$, $\epsilon$ and $\theta$.
\begin{proposition}
\label{prop:trans.geom.hom}Consider a geometric target density as
in \eqref{eq:ex.geom.density} and a proposal density as in \eqref{eq:ex.non.sym.prop}.
In addition, let the weights when $N=1$ be given by \eqref{eq:ex.Blli.weights}.
Then, for some $b$, $\varepsilon$ and $\theta$ the chain generated
by the noisy kernel $\tilde{P}_{N=1}$ is transient.
\end{proposition}
In contrast, since the weights are uniformly bounded by $b$, the pseudo-marginal
chain inherits geometric ergodicity for any $\theta$, $b$ and $\epsilon$. The left plot in \Cref{fig:discrete.examples}
shows an example. We will discuss the behaviour of this example as $N$ increases in \Cref{sub:Remarks.results} .

\subsection{Transient noisy chain with non-homogeneous weights\label{sub:Trans.nonhom.weights}}

\begin{figure}
\includegraphics[width=0.96\linewidth]{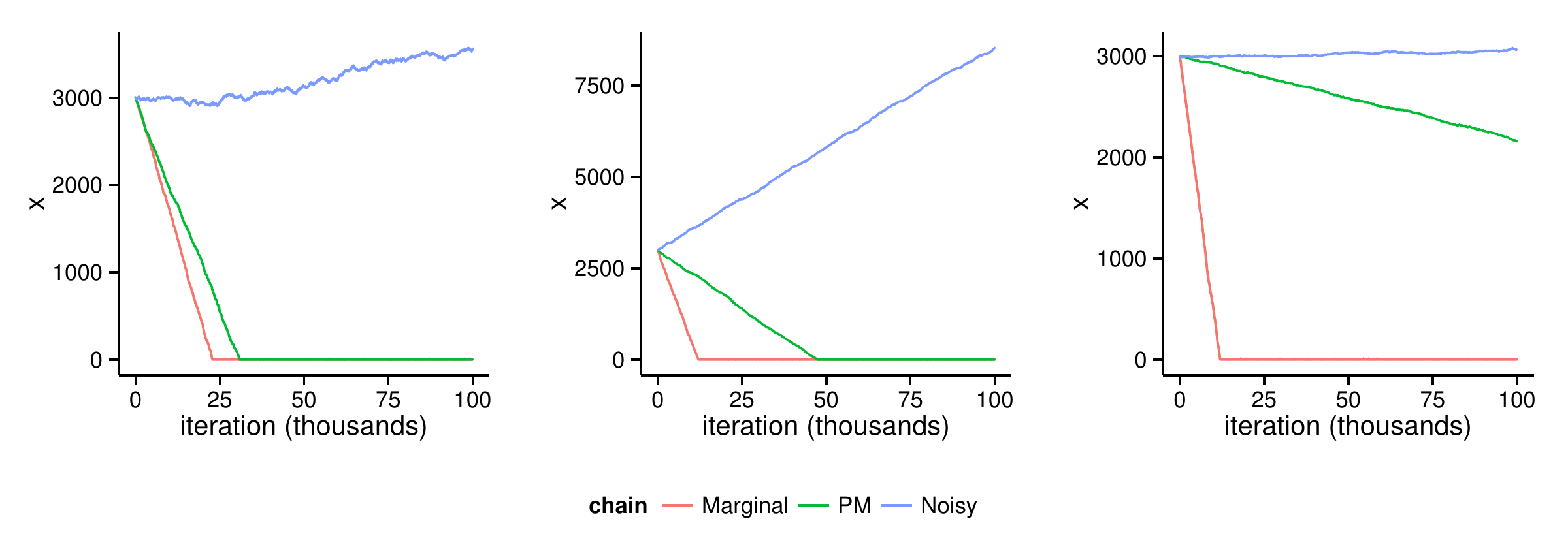}
\caption{Runs of the marginal, pseudo-marginal and noisy chains. Left plot shows example
in \Cref{sub:Trans.hom.weights}, where $\theta=0.75$, $\varepsilon=2-\sqrt{3}$
and $b=2\varepsilon\frac{\theta}{1-\theta}$. Central and right plots show example in \Cref{sub:Trans.nonhom.weights}, where $\theta=0.5$ and $\theta=0.25$ respectively, with $\varepsilon_{m}=m^{-(3-m\Mod{3})}$ and $b=3+\left(\frac{1-\theta}{\theta}\right)^{3}$.}
\label{fig:discrete.examples}
\end{figure}

One could argue that the transient behaviour of the previous example
is related to the large value of $\theta$ in the proposal distribution. However, as shown here, for any value of $\theta \in (0,1)$ one can construct weights satisfying \eqref{eq:weights.general} for which the noisy chain is transient. With the same assumptions as in the example in 
\Cref{sub:Trans.hom.weights}, except that now the distribution
of weights is not homogeneous but given by 
\begin{align}
\begin{split}W_{m,1} &= (b-\varepsilon_{m})Ber(s_{m})+\varepsilon_{m}, \quad \text{for }b>1\text{ and }\varepsilon_{m}=m^{-(3-(m\Mod{3}))},
\end{split}
\label{eq:ex.Blli.weights.inhom}
\end{align}
the noisy chain will be transient for $b$ large enough. The proof can be found in \Cref{sec:Proofs}.
\begin{proposition}
\label{prop:trans.geom.inhom}Consider a geometric target density
as in \eqref{eq:ex.geom.density} and a proposal density as in
\eqref{eq:ex.non.sym.prop}. In addition, let the weights when
$N=1$ be given by \eqref{eq:ex.Blli.weights.inhom}. Then, for any $\theta \in (0,1)$ there exists
some $b>1$ such that the chain generated by the noisy
kernel $\tilde{P}_{N=1}$ is transient.
\end{proposition}
The reason for this becomes apparent when looking at the behaviour
of the ratios of weights. Even though $\varepsilon_{m}\rightarrow0$
as $m\rightarrow\infty$, the non-monotonic behaviour of the sequence
implies
\begin{align*}
\frac{\varepsilon_{m-1}}{\varepsilon_{m}}\in & \begin{cases}
\begin{array}{c}
O\left(m^{2}\right)\\
O\left(m^{-1}\right)
\end{array}\quad\text{if} & \begin{array}{c}
m\Mod{3}=0,\\
m\Mod{3}\in \{1,2\},
\end{array}\end{cases}
\end{align*}
and
\begin{align*}
\frac{\varepsilon_{m+1}}{\varepsilon_{m}}\in & \begin{cases}
\begin{array}{c}
O\left(m^{-2}\right)\\
O\left(m\right)
\end{array}\quad\text{if} & \begin{array}{c}
m\Mod{3}=2,\\
m\Mod{3}\in \{0,1\}.
\end{array}\end{cases}
\end{align*}
Hence, the ratio of the weights can become arbitrarily large or arbitrarily
close to zero with a non-negligible probability. This allows the algorithm
to accept moves to the right more often, if $m$ is large enough.
Once again, the pseudo-marginal chain inherits the geometrically ergodic
property from the marginal. See the central and right plots of \Cref{fig:discrete.examples}
for two examples using different proposals. Again, we will come back to this example in \Cref{sub:Remarks.results}, where we look at the behaviour of the associated noisy chain as $N$ increases.

\section{Inheritance of ergodic properties\label{sec:Inheritance}}

The inheritance of various ergodic properties of the marginal chain by pseudo-marginal Markov chains has been established using techniques that are powerful but suitable only for reversible Markov chains \citep[see, e.g.][]{AndrieuNVihola_2015}. Since the noisy Markov chains treated here can be non-reversible, a suitable tool for establishing geometric ergodicity is the use of Foster--Lyapunov functions, via geometric drift towards a small set.
\begin{definition}[Small set]
 Let $P$ be the transition kernel of a Markov chain $\mathbf{\Phi}$. A subset
$C\subseteq\mathcal{X}$ is small if there exists a positive integer
$n_{0}$, $\varepsilon>0$ and a probability measure $\nu(\cdot)$
on $\left(\mathcal{X},\mathcal{B}(\mathcal{X})\right)$ such that
the following minorisation condition holds
\begin{align}
P^{n_{0}}(x,\cdot) &\geq \varepsilon\nu(\cdot),\qquad\text{for }x\in C.\label{eq:minor.condn}
\end{align}
\end{definition}

The following theorem, which is immediate from combining \citet[][Proposition~2.1]{RobertsNRosenthal_1997} and \citet[][Theorem~15.0.1]{MeynNTweedie_2009}, establishes the equivalence between geometric ergodicity
and a geometric drift condition. For any kernel $K:\mathcal{X}\times\mathcal{B}(\mathcal{X})\rightarrow[0,1]$,
let
\begin{align*}
KV & (x):=\int_{\mathcal{X}}K(x,dz)V(z).
\end{align*}

\begin{theorem}
\label{thm:GE.equiv}Suppose that $\Phi$ is a $\phi$-irreducible
and aperiodic Markov chain with transition kernel $P$ and invariant
distribution $\pi$. Then, the following statements are equivalent:
\renewcommand{\labelenumi}{\roman{enumi}.}
\begin{enumerate}
\item There exists a small set $C$, constants $\lambda<1$ and $b<\infty$, and a function $V\geq1$ finite for some $x_0\in\mathcal{X}$ satisfying the geometric drift condition
\begin{align}\label{eq:geom.drift.cond}
PV(x) &\leq \lambda V(x)+b\mathds{1}_{\left\{ x\in C\right\} },\quad \text{for }x\in\mathcal{X}.
\end{align}
\item The chain is $\pi$-a.e. geometrically ergodic, meaning that for $\pi$-a.e. $x\in\mathcal{X}$ it satisfies \eqref{eq:geom.ergo.condn} for some $V\geq1$ (which can be taken as in $(i)$) and constants $\tau<1$, $R<\infty$.
\end{enumerate}
\end{theorem}
From this point onwards, it is assumed that the marginal and noisy
chains are $\phi$-irreducible and aperiodic. In addition, for many
of the following results, it is required that

\begin{verse}
\textlabel{(P1)}\label{assm:geom.ergo.marginal} The marginal chain is
geometrically ergodic, implying its kernel $P$ satisfies the geometric
drift condition in \eqref{eq:geom.drift.cond} for some constants $\lambda<1$ and
$b<\infty$, some function $V\ge1$ and a small set $C\subseteq\mathcal{X}$.
\end{verse}

\subsection{Conditions involving a negative moment}

From the examples of the previous section, it is clear that the weights plays a fundamental role in the behaviour
of the noisy chain. The following theorem states that the noisy chain
will inherit geometric ergodicity from the marginal
under some conditions on the weights involving a strengthened version
of the Law of Large Numbers and convergence of negative moments.

\begin{verse}
\textlabel{(W1)}\label{assm:sup.WLLN} For any $\delta>0$, the weights
$\left\{ W_{x,N}\right\} _{x,N}$ satisfy
\begin{align*}
\lim_{N\rightarrow\infty}\sup_{x\in\mathcal{X}}\mathbb{P}_{Q_{x,N}}\left[\big\vert W_{x,N}-1\big\vert\geq\delta\right]= 0.
\end{align*}
\end{verse}

\begin{verse}
\textlabel{(W2)}\label{assm:sup.inv.moment} The weights $\left\{ W_{x,N}\right\} _{x,N}$
satisfy
\begin{align*}
\lim_{N\rightarrow\infty}\sup_{x\in\mathcal{X}}\mathbb{E}_{Q_{x,N}}\left[W_{x,N}^{-1}\right]=  1.
\end{align*}
\end{verse}

\begin{theorem}
\label{thm:strong.result}Assume \ref{assm:geom.ergo.marginal},
\ref{assm:sup.WLLN} and \ref{assm:sup.inv.moment}. Then,
there exists $N_{0}\in\mathbb{N}^{+}$ such that for all $N\geq N_{0}$,
the noisy chain with transition kernel $\tilde{P}_{N}$ is geometrically
ergodic.
\end{theorem}
The above result is obtained by controlling the dissimilarity of the
marginal and noisy kernels. This is done by looking at the corresponding
rejection and acceptance probabilities. The proofs of the following lemmas appear in \Cref{sec:Proofs}.

\begin{lemma}
\label{lem:sup.probs}For any $\delta>0$
\begin{align*}
\mathbb{P}_{Q_{z,N} \otimes Q_{x,N}}\left[\frac{W_{z,N}}{W_{x,N}}\leq 1-\delta\right] &\leq 2\sup_{x\in\mathcal{X}}\mathbb{P}_{Q_{x,N}}\left[\Big\vert W_{x,N}-1\Big\vert\geq\frac{\delta}{2}\right].
\end{align*}
\end{lemma}

\begin{lemma}
\label{lem:rejec.probs}Let $\rho(x)$ and $\tilde{\rho}_{N}(x)$
be the rejection probabilities as defined in \eqref{eq:alpha_rho}
and \eqref{eq:rho_tilde} respectively. Then, for any $\delta>0$
\begin{align*}
\tilde{\rho}_{N}(x)-\rho(x)\leq \delta+2\sup_{x\in\mathcal{X}}\mathbb{P}_{Q_{x,N}}\left[\big\vert W_{x,N}-1\big\vert\geq\frac{\delta}{2}\right].
\end{align*}
\end{lemma}

\begin{lemma}
\label{lem:accep.probs.prop}Let $\alpha(x,y)$ and $\tilde{\alpha}_{N}(x,y)$
be the acceptance probabilities as defined in \eqref{eq:alpha_rho}
and \eqref{eq:alpha_tilde} respectively. Then,
\begin{align*}
\tilde{\alpha}_{N}(x,y) \leq \alpha(x,y)\mathbb{E}_{Q_{x,N}}\left[W_{x,N}^{-1}\right].
\end{align*}
\end{lemma}
Notice that \ref{assm:sup.WLLN} and \ref{assm:sup.inv.moment}
allow control on the bounds in the above lemmas. While \Cref{lem:rejec.probs}
provides a bound for the difference of the rejection probabilities,
\Cref{lem:accep.probs.prop} gives one for the ratio of the acceptance
probabilities. The proof of \Cref{thm:strong.result} is now
presented.

\begin{prooflabel}[\Cref{thm:strong.result}]
Since the marginal chain $P$ is geometrically ergodic, it satisfies
the geometric drift condition in \eqref{eq:geom.drift.cond} for some
$\lambda<1$, $b<\infty$, some function $V\geq1$ and a small set
$C\subseteq\mathcal{X}$. Now, using the above lemmas 
\begin{align*}
\tilde{P}_{N}V(x) -PV(x) &= \int_{\mathcal{X}}q(x,dz)\left(\tilde{\alpha}_{N}(x,z)-\alpha(x,z)\right)V(z) +V(x)\left(\tilde{\rho}_{N}(x)-\rho(x)\right)\\
&\leq \left(\sup_{x\in\mathcal{X}}\mathbb{E}\left[W_{x,N}^{-1}\right]-1\right)PV(x) +\left(\delta+2\sup_{x\in\mathcal{X}}\mathbb{P}\left[\big\vert W_{x,N}-1\big\vert\geq\frac{\delta}{2}\right]\right)V(x).
\end{align*}

By \ref{assm:sup.WLLN} and \ref{assm:sup.inv.moment}, for any $\varepsilon$, $\delta>0$ there exists $N_{0}\in\mathbb{N}^{+}$ such that 
\begin{align*}
\sup_{x\in\mathcal{X}}\mathbb{P}\left[\big\vert W_{x,N}-1\big\vert\geq\frac{\delta}{2}\right]<\frac{\varepsilon}{4} \quad\text{and} \quad \sup_{x\in\mathcal{X}}\mathbb{E}\left[W_{x,N}^{-1}\right]-1<\varepsilon,
\end{align*}
whenever $N\geq N_{0}$, implying
\begin{align*}
\tilde{P}_{N}V(x) &\leq PV(x)+\varepsilon PV(x)+\left(\delta+\frac{\varepsilon}{2}\right)V(x)\\
&\leq \lambda\left(1+\delta+\frac{\varepsilon}{2}\right)V(x)+b\left(1+\varepsilon\right)\mathds{1}_{\left\{ x\in C\right\} }.
\end{align*}
Taking $\delta=\frac{\varepsilon}{2}$ and $\varepsilon\in\left(0,\frac{1-\lambda}{\lambda}\right)$,
the noisy chain $\tilde{P}_{N}$ also satisfies a geometric drift
condition for the same function $V$ and small set $C$, completing
the proof.
\end{prooflabel}

\subsection{Conditions on the proposal distribution}

In this subsection a different bound for the acceptance probabilities
is provided, which allows dropping assumption \ref{assm:sup.inv.moment}
but imposes a different one on the proposal $q$ instead.

\begin{verse}
\textlabel{(P1{*})}\label{assm:qV.less.KV} \ref{assm:geom.ergo.marginal} holds
and for the same drift function $V$ in  \ref{assm:geom.ergo.marginal} there exists $K<\infty$ such that the proposal kernel
$q$ satisfies
\begin{align*}
qV(x) &\leq KV(x),\quad\text{for }x\in\mathcal{X}.
\end{align*}
\end{verse}

\begin{theorem}
\label{thm:weak.result}Assume \ref{assm:qV.less.KV} and \ref{assm:sup.WLLN}.
Then, there exists $N_{0}\in\mathbb{N}^{+}$ such that for all $N\geq N_{0}$,
the noisy chain with transition kernel $\tilde{P}_{N}$ is geometrically
ergodic.
\end{theorem}

In order to prove \Cref{thm:weak.result} the following lemma is required. Its proof can be found in \Cref{sec:Proofs}. In contrast with \Cref{lem:accep.probs.prop}, this lemma provides a bound for the additive difference of the noisy and marginal acceptance probabilities.
\begin{lemma}
\label{lem:accept.probs.linear}Let $\alpha(x,y)$ and $\tilde{\alpha}_{N}(x,y)$
be the acceptance probabilities as defined in \eqref{eq:alpha_rho}
and \eqref{eq:alpha_tilde}, respectively. Then, for any $\eta>0$
\begin{align*}
\tilde{\alpha}_{N}(x,y) -\alpha(x,y) &\leq \eta+2\sup_{x\in\mathcal{X}}\mathbb{P}_{Q_{x,N}}\left[\Big\vert W_{x,N}-1\Big\vert\geq\frac{\eta}{2\left(1+\eta\right)}\right].
\end{align*}
\end{lemma}

\begin{prooflabel}[\Cref{thm:weak.result}]
Using \Cref{lem:rejec.probs} and \Cref{lem:accept.probs.linear}
with $\eta=\delta$ 
\begin{align*}
\tilde{P}_{N}&V(x)-PV(x)\\
&= \int_{\mathcal{X}}q(x,dz)\left(\tilde{\alpha}_{N}(x,z)-\alpha(x,z)\right)V(z)+V(x)\left(\tilde{\rho}_{N}(x)-\rho(x)\right)\\
&\leq \left(\delta+2\sup_{x\in\mathcal{X}}\mathbb{P}\left[\Big\vert W_{x,N}-1\Big\vert\geq\frac{\delta}{2\left(1+\delta\right)}\right]\right)qV(x) +\left(\delta+2\sup_{x\in\mathcal{X}}\mathbb{P}\left[\Big\vert W_{x,N}-1\Big\vert\geq\frac{\delta}{2}\right]\right)V(x)\\
&\leq \left(\delta+2\sup_{x\in\mathcal{X}}\mathbb{P}\left[\Big\vert W_{x,N}-1\Big\vert\geq\frac{\delta}{2\left(1+\delta\right)}\right]\right) \left(qV(x)+V(x)\right).
\end{align*}
By \ref{assm:sup.WLLN}, there exists $N_{0}\in\mathbb{N}^{+}$ such that 
\begin{align*}
\sup_{x\in\mathcal{X}}\mathbb{P}\left[\Big\vert W_{x,N}-1\Big\vert\geq\frac{\delta}{2\left(1+\delta\right)}\right] & <\frac{\varepsilon}{4},
\end{align*}
whenever $N\geq N_{0}$. This implies
\begin{align*}
\tilde{P}_{N}V(x) &\leq PV(x)+\left(\delta+\frac{\varepsilon}{2}\right)\left(qV(x)+V(x)\right),
\end{align*}
and using \ref{assm:qV.less.KV}
\begin{align*}
\tilde{P}_{N}V(x) &\leq \left(\lambda+\left(\delta+\frac{\varepsilon}{2}\right)\left(K+1\right)\right)V(x)+b\mathds{1}_{\left\{ x\in C\right\} }.
\end{align*}
Taking $\delta=\frac{\varepsilon}{2}$ and $\varepsilon\in\left(0,\frac{1-\lambda}{1+K}\right)$,
the noisy chain $\tilde{P}_{N}$ also satisfies a geometric drift
condition for the same function $V$ and small set $C$, completing
the proof.
\end{prooflabel}

In general, assumption \ref{assm:qV.less.KV} may be difficult to
verify as one must identify a particular function $V$, but it is easily satisfied when restricting to log-Lipschitz
targets and when using a random walk proposal of the form 
\begin{align}
q(x,dy) &= q(\|y-x\|)dy,\label{eq:rw.proposal}
\end{align}
where $\|\cdot\|$ denotes the usual Euclidean distance. To see this the following assumption is required, which is a particular
case of \ref{assm:geom.ergo.marginal} and is satisfied under
some extra technical conditions \citep[see, e.g.,][]{RobertsNTweedie_1996}.

\begin{verse}
\textlabel{(P1{*}{*})}\label{assm:marginal.drift.V.as.pi} $\mathcal{X}\subseteq\mathbb{R}^{d}$. The target $\pi$ is log-Lipschitz, meaning that for some $L>0$
\begin{align*}
|\log\pi(z)-\log\pi(x)| &\leq L\|z-x\|.
\end{align*}
\ref{assm:geom.ergo.marginal} holds taking the drift function $V=\pi^{-s}$, for any $s\in(0,1)$. The proposal $q$ is a random walk as in \eqref{eq:rw.proposal} satisfying
\begin{align*}
\int_{\mathbb{R}^{d}}\exp\left\{ a\|u\|\right\} q(\|u\|)du &< \infty,
\end{align*}
for some $a>0$.
\end{verse}

See \Cref{sec:Proofs} for a proof of the following proposition.
\begin{proposition}
\label{prop:log.Lipschitz}Assume \ref{assm:marginal.drift.V.as.pi} and
\ref{assm:sup.WLLN}. Then, \ref{assm:qV.less.KV} holds.
\end{proposition}

\subsection{Conditions for arithmetic averages}

In the particular setting where the weights are given by \eqref{eq:wghts.arithm.avg},
sufficient conditions on these can be obtained to ensure geometric
ergodicity is inherited by the noisy chain. For the simple case where the weights are homogeneous with respect to the state space
\ref{assm:sup.WLLN} is automatically satisfied. In order to attain \ref{assm:sup.inv.moment}, the existence of a negative moment for a single weight is required. See \Cref{sec:Proofs} for a proof of the following result.
\begin{proposition}\label{prop:convergence.fatous}
Assume weights as in \eqref{eq:wghts.arithm.avg}. If $\mathbb{E}_{Q_x}\left[W_x^{-1}\right]<\infty$ then
\begin{align}\label{eq:result.fatous}
\lim_{N\rightarrow\infty}\mathbb{E}_{Q_{x,N}}\left[W_{x,N}^{-1}\right]= 1.
\end{align}
\end{proposition}

For homogeneous weights, \eqref{eq:result.fatous} implies \ref{assm:sup.inv.moment}. When the weights are not homogeneous, stronger conditions are needed for \ref{assm:sup.WLLN} and \ref{assm:sup.inv.moment} to be satisfied. An appropriate first assumption is that the weights are uniformly integrable.
\begin{verse}
\textlabel{(W3)}\label{assm:unif.integ} The weights $\left\{ W_{x}\right\} _{x}$
satisfy
\begin{align*}
\lim_{K\rightarrow\infty}\sup_{x\in\mathcal{X}}\mathbb{E}_{Q_{x}}\left[W_{x}\mathds{1}_{\{W_{x}>K\}}\right] & =0.
\end{align*}
\end{verse}
The second condition imposes an additional assumption on the distribution
of the weights $\left\{ W_{x}\right\} _{x}$ near $0$.
\begin{verse}
\textlabel{(W4)}\label{assm:unif.bdd.densities} There exists $\gamma\in(0,1)$ and constants $M<\infty$, $\beta>0$ such that for $w\in (0,\gamma)$ the weights $\left\{ W_{x}\right\} _{x}$ satisfy
\begin{align*}
\sup_{x\in\mathcal{X}} \mathbb{P}_{Q_x} \left[ W_x \leq w \right] & \leq M w^{\beta}.
\end{align*}
\end{verse}
These new conditions ensure \ref{assm:sup.WLLN} and \ref{assm:sup.inv.moment} are satisfied.

\begin{proposition}\label{prop:equivalence.conditions}
For weights as in \eqref{eq:wghts.arithm.avg}, 
\renewcommand{\labelenumi}{\roman{enumi}.}
\begin{enumerate}
\item \ref{assm:unif.integ} implies \ref{assm:sup.WLLN};
\item \ref{assm:sup.WLLN} and \ref{assm:unif.bdd.densities} imply \ref{assm:sup.inv.moment}.
\end{enumerate}
\end{proposition}

The following corollary is obtained as an immediate consequence of the above proposition, \Cref{thm:strong.result} and \Cref{thm:weak.result}.
\begin{corollary}\label{cor:strong.and.weak.result}
Let the weights be as in \eqref{eq:wghts.arithm.avg}. Assume \ref{assm:unif.integ} and either
\renewcommand{\labelenumi}{\roman{enumi}.}
\begin{enumerate}
\item \ref{assm:geom.ergo.marginal} and \ref{assm:unif.bdd.densities};
\item \ref{assm:qV.less.KV}.
\end{enumerate}
Then, there exists $N_{0}\in\mathbb{N}^{+}$ such that for all $N\geq N_{0}$, the noisy
chain with transition kernel $\tilde{P}_{N}$ is geometrically ergodic.
\end{corollary}

The proof of \Cref{prop:equivalence.conditions} follows the statement of \Cref{lem:sums.rvs}, whose proof can be found in \Cref{sec:Proofs}. This lemma allows us to characterise the distribution of $W_{x,N}$ near 0 assuming \ref{assm:unif.bdd.densities} and also provides conditions for the existence and convergence of negative moments.

\begin{lemma}\label{lem:sums.rvs}
Let $\gamma\in(0,1)$ and $p>0$.
\renewcommand{\labelenumi}{\roman{enumi}.}
\begin{enumerate}
\item Suppose $Z$ is a positive random variable, and assume that for $z\in(0,\gamma)$
\begin{align*}
\mathbb{P}\left[ Z \leq z \right] &\leq Mz^{\alpha},\quad\text{where }\alpha>p,M<\infty.
\end{align*}
Then,
\begin{align*}
\mathbb{E}\left[ Z^{-p} \right] \leq \frac{1}{\gamma^p} + pM\frac{\gamma^{\alpha-p}}{\alpha-p}.
\end{align*}

\item Suppose $\left\{ Z_{i}\right\} _{i=1}^{N}$ is a collection of positive and independent random variables, and assume that for each $i\in\left\{ 1,\dots,N\right\}$ and $z\in(0,\gamma)$
\begin{align*}
\mathbb{P}\left[ Z_{i} \leq z \right] &\leq M_{i} z^{\alpha_{i}},\quad\text{where }\alpha_{i}>0,M_{i}<\infty.
\end{align*}
Then, for $z\in(0,\gamma)$
\begin{align*}
\mathbb{P}\left[\sum_{i=1}^{N}Z_{i}\leq z\right] \leq \prod_{i=1}^N M_i z^{\sum_{i=1}^{N}\alpha_{i}}.
\end{align*}

\item Let the weights be as in \eqref{eq:wghts.arithm.avg}. If for some $N_0\in\mathbb{N}^+$
\begin{align*}
\mathbb{E}_{Q_{x,N_0}}\left[ W_{x,N_0}^{-p} \right] < \infty,
\end{align*}
then for any $N\geq N_0$
\begin{align*}
\mathbb{E}_{Q_{x,N+1}}\left[ W_{x,N+1}^{-p} \right] \leq \mathbb{E}_{Q_{x,N}}\left[ W_{x,N}^{-p} \right].
\end{align*}

\item Assume \ref{assm:sup.WLLN} and let $g:\mathbb{R}^{+}\rightarrow\mathbb{R}$ be a function
that is continuous at 1 and bounded on the interval $[\gamma,\infty)$. Then
\begin{align*}
\lim_{N\rightarrow\infty}\sup_{x\in\mathcal{X}}\mathbb{E}_{Q_{x,N}}\left[|g\left(W_{x,N}\right)-g\left(1\right)|\mathds{1}_{W_{x,N}\geq\gamma}\right] = 0.
\end{align*}
\end{enumerate}
\end{lemma}

\begin{prooflabel}[\Cref{prop:equivalence.conditions}]
Part $(i)$ is a consequence of \citet[][Theorem~1]{Chandra_1989}. Assuming \ref{assm:unif.integ}, it implies
\begin{align*}
\lim_{N\rightarrow\infty}\sup_{x\in\mathcal{X}}\mathbb{E}\big\vert W_{x,N}-1\big\vert = 0.
\end{align*}
By Markov's inequality 
\begin{align*}
\mathbb{E}\big\vert W_{x,N}-1\big\vert \geq \delta \mathbb{P}\left[ \big\vert W_{x,N}-1\big\vert \geq \delta \right],
\end{align*}
and the result follows.

To prove $(ii)$, assume \ref{assm:unif.bdd.densities} and by part $(ii)$ of \Cref{lem:sums.rvs}, for $w\in(0,\gamma)$
\begin{align*}
\mathbb{P}\left[ N W_{x,N} \leq w \right]\leq M^N w^{N\beta}.
\end{align*}
Take $p>1$ and define $N_0:=\lfloor \frac{p}{\beta} \rfloor+1$, then using part $(i)$ of \Cref{lem:sums.rvs} if $N\geq N_0$
\begin{align*}
\sup_{x\in\mathcal{X}}\mathbb{E}\left[ W_{x,N}^{-p} \right] \leq \frac{N}{\gamma^p} + pNM^N \frac{\gamma^{N\beta-p}}{N\beta-p}.
\end{align*}
Hence, by H\"older's inequality
\begin{align*}
\mathbb{E} \left[\big\vert W_{x,N}^{-1}-1\big\vert\mathds{1}_{W_{x,N}\in(0,\gamma)}\right] &\leq \mathbb{E}\left[ W_{x,N}^{-1} \mathds{1}_{W_{x,N}\in(0,\gamma)}\right]\\
&\leq \left(\mathbb{E}\left[ W_{x,N}^{-p} \right]\right)^\frac{1}{p} \left( \mathbb{P}\left[W_{x,N} < \gamma \right]\right)^{\frac{p-1}{p}},
\end{align*}
and applying part $(iii)$ of \Cref{lem:sums.rvs}, for $N\geq N_0$
\begin{align*}
\mathbb{E} \left[\big\vert W_{x,N}^{-1}-1\big\vert\mathds{1}_{W_{x,N}\in(0,\gamma)}\right] &\leq \left(\mathbb{E}\left[ W_{x,N_0}^{-p} \right]\right)^\frac{1}{p} \left( \mathbb{P}\left[W_{x,N} < \gamma \right]\right)^{\frac{p-1}{p}}.
\end{align*}
Therefore, 
\begin{align*}
\sup_{x\in\mathcal{X}}\mathbb{E} \left[\big\vert W_{x,N}^{-1}-1\big\vert\mathds{1}_{W_{x,N}\in(0,\gamma)}\right] &\leq \left(\sup_{x\in\mathcal{X}} \mathbb{E}\left[ W_{x,N_0}^{-p} \right]\right)^\frac{1}{p} \left( \sup_{x\in\mathcal{X}} \mathbb{P}\left[W_{x,N} < \gamma \right]\right)^{\frac{p-1}{p}}.
\end{align*}
Since $\gamma<1$ and by \ref{assm:sup.WLLN}
\begin{align*}
\lim_{N\rightarrow\infty}\sup_{x\in\mathcal{X}} \mathbb{P}\left[W_{x,N} < \gamma \right]=0,
\end{align*}
implying
\begin{align}\label{eq:first.half}
\lim_{N\rightarrow\infty}\sup_{x\in\mathcal{X}}\mathbb{E} & \left[\big\vert W_{x,N}^{-1}-1\big\vert\mathds{1}_{W_{x,N}\in(0,\gamma)}\right]=0.
\end{align}
Now, for fixed $\gamma\in(0,1)$ the function $g(x)=x^{-1}$
is bounded and continuous on $[\gamma,\infty)$, implying by part $(iv)$ of \Cref{lem:sums.rvs}
\begin{align}\label{eq:second.half}
\lim_{N\rightarrow\infty}\sup_{x\in\mathcal{X}}\mathbb{E}\left[\big\vert W_{x,N}^{-1}-1\big\vert\mathds{1}_{W_{x,N}\in[\gamma,\infty)}\right] = 0.
\end{align}
Finally, using \eqref{eq:first.half} and \eqref{eq:second.half}
\begin{align*}
\lim_{N\rightarrow\infty}\sup_{x\in\mathcal{X}} & \mathbb{E}\big\vert W_{x,N}^{-1}-1\big\vert = 0,
\end{align*}
and by the triangle inequality
\begin{align*}
\sup_{x\in\mathcal{X}}\mathbb{E}\big\vert W_{x,N}^{-1}-1\big\vert \geq \sup_{x\in\mathcal{X}}\mathbb{E}\left[W_{x,N}^{-1}\right]-1,
\end{align*}
the result follows.
\end{prooflabel}

\subsection{Remarks on results\label{sub:Remarks.results}}

Equipped with these results, we return to the examples in
\Cref{sub:Trans.hom.weights} and \Cref{sub:Trans.nonhom.weights}.
Even though the noisy chain can be transient in these examples, the behaviour
is quite different when considering weights that are arithmetic averages of the form
in \eqref{eq:wghts.arithm.avg}. Since in both examples the weights
are uniformly bounded by the constant $b$, they immediately satisfy
\ref{assm:sup.WLLN}. Additionally, by \Cref{prop:convergence.fatous},
condition \ref{assm:sup.inv.moment} is satisfied for the example
in \Cref{sub:Trans.hom.weights}. This is not the case for
example in \Cref{sub:Trans.nonhom.weights}, but condition
\ref{assm:qV.less.KV} is satisfied by taking $V=\pi^{-\frac{1}{2}}$.
Therefore, applying \Cref{thm:strong.result} and \Cref{thm:weak.result}
to examples in \Cref{sub:Trans.hom.weights} and in
\Cref{sub:Trans.nonhom.weights} respectively, as $N$ increases
the corresponding chains will go from being transient to
geometrically ergodic.

Despite conditions \ref{assm:sup.WLLN} and \ref{assm:sup.inv.moment} guaranteeing the inheritance of geometric ergodicity
for the noisy chain, they are not necessary. Consider a modification of the example in \Cref{sub:Trans.hom.weights}, where the weights are given by 
\begin{align*}
W_{m,1} &= (b_{m}-\varepsilon_{m})Ber(s_{m})+\varepsilon_{m}, \quad \text{where }b_{m}>1\text{ and }\varepsilon_{m}\in(0,1]\text{ for all }m\geq1.
\end{align*}
Again, there exists a relationship between the variables $b_{m}$, $\varepsilon_{m}$
and $s_{m}$ for ensuring the expectation of the weights is equal
to one. Let $Bin\left(N,s\right)$ denote a binomial distribution
of parameters $N\in\mathbb{N}^{+}$ and $s\in(0,1)$. Then, in the
arithmetic average context, $W_{m,N}$ becomes
\begin{align}
\begin{split}W_{m,N} &= \frac{\left(b_{m}-\varepsilon_{m}\right)}{N}Bin\left(N,s_{m}\right)+\varepsilon_{m}, \quad \text{where }b_{m}>1\text{ and }\varepsilon_{m}\in(0,1]\text{ for all }m\geq1.
\end{split}
\label{eq:weight.binom}
\end{align}
For particular choices of the sequences $\left\{ b_{m}\right\} _{m\in\mathbb{N}^{+}}$
and $\left\{ \varepsilon_{m}\right\} _{m\in\mathbb{N}^{+}}$, the resulting noisy chain can
be geometrically ergodic for all $N\geq1$, even though neither 
\ref{assm:sup.WLLN} nor \ref{assm:sup.inv.moment} hold.

\begin{proposition}\label{prop:GE.geom.inhom.allN}
Consider a geometric target density
as in \eqref{eq:ex.geom.density} and a proposal density as in
\eqref{eq:ex.non.sym.prop}. In addition, let the weights be as
in \eqref{eq:weight.binom} with $b_{m}\rightarrow\infty$, $\varepsilon_{m}\rightarrow0$
as $m\rightarrow\infty$ and
\begin{align*}
\lim_{m\rightarrow\infty}\frac{\varepsilon_{m-1}}{\varepsilon_{m}} = l,\quad\text{where }l\in\mathbb{R}^{+}\cup\left\{ +\infty\right\} .
\end{align*}
Then, the chain generated by the noisy kernel $\tilde{P}_{N}$ is
geometrically ergodic for any $N\in\mathbb{N}^{+}$.
\end{proposition}

Finally, in many of the previous examples, increasing the value of
$N$ seems to improve the ergodic properties of the noisy chain. However,
the geometric ergodicity property is not always inherited, no matter
how large $N$ is taken. 
The following proposition shows an example rather similar to \Cref{prop:GE.geom.inhom.allN}, but in which the ratio $\frac{\varepsilon_{m-1}}{\varepsilon_{m}}$ does not converge as $m\rightarrow\infty$.
\begin{proposition}
\label{prop:trans.geom.inhom.allN}Consider a geometric target density
as in \eqref{eq:ex.geom.density} and a proposal density as in
\eqref{eq:ex.non.sym.prop}. In addition, let the weights be as
in \eqref{eq:weight.binom} with $b_{m}=m$ and
\begin{align*}
\varepsilon_{m} = m^{-(3-(m\Mod{3}))}.
\end{align*}
Then, the chain generated by the noisy kernel $\tilde{P}_{N}$ is
transient for any $N\in\mathbb{N}^{+}$.
\end{proposition}

\section{Convergence of the noisy invariant distribution\label{sec:Invariants}}

So far the only concern has been whether the noisy chain inherits
the geometric ergodicity property from the marginal chain. As an immediate
consequence, geometric ergodicity guarantees the existence of an invariant
probability distribution $\tilde{\pi}_{N}$ for $\tilde{P}_{N}$,
provided $N$ is large enough. In addition, using the same conditions from \Cref{sec:Inheritance}, we can characterise and in some cases quantify the convergence in total variation of $\tilde{\pi}_{N}$ towards the desired target $\pi$, as $N\rightarrow\infty	$. 

\subsection{Convergence in total variation}

The following definition, taken from \citet{RobertsETal_1998}, characterises a class of kernels satisfying a geometric drift condition as in \eqref{eq:geom.drift.cond} for the same $V$, $C$, $\lambda$ and $b$.

\begin{definition}[Simultaneous geometric ergodicity]
\label{defn:sim.geom.erg}A class of Markov chain kernels $\left\{ P_{k}\right\} _{k\in\mathcal{K}}$
is simultaneously geometrically ergodic if there exists a class of
probability measures $\left\{ \nu_{k}\right\} _{k\in\mathcal{K}}$,
a measurable set $C\subseteq\mathcal{X}$, a real valued measurable
function $V\geq1$, a positive integer $n_{0}$ and positive constants
$\varepsilon$, $\lambda$, $b$ such that for each $k\in\mathcal{K}$:
\renewcommand{\labelenumi}{\roman{enumi}.}
\begin{enumerate}
\item $C$ is small for $P_{k}$, with $P_{k}^{n_{0}}(x,\cdot)\geq\varepsilon\nu_{k}(\cdot)$
for all $x\in C$;
\item the chain $P_{k}$ satisfies the geometric drift condition in \eqref{eq:geom.drift.cond}
with drift function $V$, small set $C$ and constants $\lambda$ and $b$.
\end{enumerate}
\end{definition}

Provided $N$ is large, the noisy kernels $\{ \tilde{P}_{N+k}\}_{k\geq 0}$ together with the marginal $P$ will be simultaneous geometrically ergodic. This
will allow the use of coupling arguments for ensuring $\tilde{\pi}_{N}$
and $\pi$ get arbitrarily close in total variation. The main additional assumption is

\begin{verse}
\textlabel{(P2)}\label{assm:minor.condn.subkernel} For some $\varepsilon>0$,
some probability measure $\nu(\cdot)$ on $\left(\mathcal{X},\mathcal{B}(\mathcal{X})\right)$
and some subset $C\subseteq\mathcal{X}$, the marginal acceptance
probability $\alpha$ and the proposal kernel $q$ satisfy
\begin{align*}
\alpha(x,y)q(x,dy) \geq \varepsilon\nu(dy),\qquad\text{for }x\in C.
\end{align*}
\end{verse}

\begin{remark}
\ref{assm:minor.condn.subkernel} ensures the marginal chain satisfies the
minorisation condition in \eqref{eq:minor.condn}, purely attained
by the sub-kernel $\alpha(x,y)q(x,dy)$. This occurs under fairly mild assumptions
\citep[see, e.g.,][Theorem~2.2]{RobertsNTweedie_1996}.\end{remark}

\begin{theorem}
\label{thm:close.invariants}Assume \ref{assm:geom.ergo.marginal},
\ref{assm:minor.condn.subkernel}, \ref{assm:sup.WLLN}
and \ref{assm:sup.inv.moment}. Alternatively, assume \ref{assm:qV.less.KV},
\ref{assm:minor.condn.subkernel} and \ref{assm:sup.WLLN}.
Then, \renewcommand{\labelenumi}{\roman{enumi}.} 
\begin{enumerate}
\item there exists $N_{0}\in\mathbb{N}^{+}$ such that the class of kernels
$\left\{ P,\tilde{P}_{N_{0}},\tilde{P}_{N_{0}+1},\dots\right\} $
is simultaneously geometrically ergodic;
\item for all $x\in\mathcal{X}$, $\lim_{N\rightarrow\infty}\|\tilde{P}_{N}(x,\cdot)-P(x,\cdot)\|_{TV}=0$;
\item $\lim_{N\rightarrow\infty}\|\tilde{\pi}_{N}(\cdot)-\pi(\cdot)\|_{TV}=0.$
\end{enumerate}
\end{theorem}

Part $(iii)$ of the above theorem is mainly a consequence of \citet[][Theorem~9]{RobertsETal_1998} when parts
$(i)$ and $(ii)$ hold. Indeed, by the triangle inequality,
\begin{align}
\begin{split}\|\tilde{\pi}_{N}(\cdot) - \pi(\cdot)\|_{TV} &\leq \|\tilde{P}_{N}^{n}(x,\cdot)-\tilde{\pi}_{N}(\cdot)\|_{TV}+\|P^{n}(x,\cdot)-\pi(\cdot)\|_{TV} +\|\tilde{P}_{N}^{n}(x,\cdot)-P^{n}(x,\cdot)\|_{TV}.
\end{split}
\label{eq:tri.ineq.invariants}
\end{align}
Provided $N\geq N_{0}$, the first two terms in \eqref{eq:tri.ineq.invariants}
can be made arbitrarily small by increasing $n$. In addition, due
to the simultaneous geometrically ergodic property, the first term
in \eqref{eq:tri.ineq.invariants} is uniformly controlled regardless
the value of $N$. Finally, using an inductive argument, part $(ii)$
implies that for all $x\in\mathcal{X}$ and all $n\in\mathbb{N}^{+}$
\begin{align*}
\lim_{N\rightarrow\infty}\|\tilde{P}_{N}^{n}(x,\cdot)-P^{n}(x,\cdot)\|_{TV} = 0.
\end{align*}

\begin{prooflabel}[\Cref{thm:close.invariants}]
From the proofs of \Cref{thm:strong.result} and \Cref{thm:weak.result}, there exists $N_{2}\in\mathbb{N}^{+}$ such
that the class of kernels $\left\{ P,\tilde{P}_{N_{2}},\tilde{P}_{N_{2}+1},\dots\right\} $
satisfies condition $(ii)$ in \Cref{defn:sim.geom.erg}
for the same function $V$, small set $C$ and constants $\lambda_{N_{2}},b_{N_{2}}$.
Respecting $(i)$, for any $\delta\in(0,1)$ 
\begin{align*}
\tilde{P}_{N} (x,A) &\geq \int_{A}\tilde{\alpha}_{N}(x,z)q(x,dz)\\
&\geq \int_{A}\mathbb{E}\left[\min\left\{ 1,\frac{W_{z,N}}{W_{x,N}}\right\} \right]\alpha(x,z)q(x,dz)\\
&\geq (1-\delta) \int_{A}\left(1-\mathbb{P}\left[\frac{W_{z,N}}{W_{x,N}}\leq1-\delta\right]\right)\alpha(x,z)q(x,dz).
\end{align*}
Then, by \Cref{lem:sup.probs}
\begin{align*}
\tilde{P}_{N}(x,A) &\geq (1-\delta)\left(1-2\sup_{x\in\mathcal{X}}\mathbb{P}\left[\Big\vert W_{x,N}-1\Big\vert\geq\frac{\delta}{2}\right]\right) \int_{A}\alpha(x,z)q(x,dz).
\end{align*}
By \ref{assm:sup.WLLN}, there exists $N_{1}\in\mathbb{N}^{+}$
such that for $N\geq N_{1}$ 
\begin{align*}
\sup_{x\in\mathcal{X}}\mathbb{P}\left[\Big\vert W_{x,N}-1\Big\vert\geq\frac{\delta}{2}\right] \leq \frac{\delta}{2},
\end{align*}
giving
\begin{align*}
\tilde{P}_{N}(x,A) \geq (1-\delta)^{2}\int_{A}\alpha(x,z)q(x,dz).
\end{align*}
Due to \ref{assm:minor.condn.subkernel},
\begin{align*}
\tilde{P}_{N}(x,A) \geq (1-\delta)^{2}\varepsilon\nu(A),\qquad\text{for }x\in C.
\end{align*}
Finally, take $N_{0}=\max\left\{ N_{1},N_{2}\right\} $ implying
$(i)$.

To prove $(ii)$ apply \Cref{lem:rejec.probs} and \Cref{lem:accept.probs.linear}
to get
\begin{align}
\sup_{A\in\mathcal{B}(\mathcal{X})}\left\{ \tilde{P}_{N}(x,A)-P(x,A)\right\} &\leq \left(\eta+2\sup_{x\in\mathcal{X}}\mathbb{P}_{Q_{x,N}}\left[\Big\vert W_{x,N}-1\Big\vert\geq\frac{\eta}{2\left(1+\eta\right)}\right]\right) \sup_{A\in\mathcal{B}(\mathcal{X})}q(x,A) \nonumber\\
 &\quad +\left(\tilde{\rho_{N}}(x)-\rho(x)\right)\sup_{A\in\mathcal{B}(\mathcal{X})}\mathds{1}_{x\in A} \nonumber\\
&\leq \left(\eta+2\sup_{x\in\mathcal{X}}\mathbb{P}_{Q_{x,N}}\left[\Big\vert W_{x,N}-1\Big\vert\geq\frac{\eta}{2\left(1+\eta\right)}\right]\right) \nonumber\\
 &\quad +\left(\delta+2\sup_{x\in\mathcal{X}}\mathbb{P}_{Q_{x,N}}\left[\big\vert W_{x,N}-1\big\vert\geq\frac{\delta}{2}\right]\right) \label{eq:aux.bound.tv}
\end{align}
Finally, taking $N\rightarrow\infty$ and by \ref{assm:sup.WLLN}
\begin{align*}
\lim_{N\rightarrow\infty}\sup_{A\in\mathcal{B}(\mathcal{X})}\left\{ \tilde{P}_{N}(x,A)-P(x,A)\right\} \leq \eta+\delta.
\end{align*}
The result follows since $\eta$ and $\delta$ can be taken arbitrarily
small.

For $(iii)$, see Theorem 9 in \citet{RobertsETal_1998} for a detailed proof.
\end{prooflabel}

\subsection{Rate of convergence}

Let $(\tilde{\Phi}_{n}^{N})_{n\geq0}$ denote the noisy chain and $\left(\Phi_{n}\right)_{n\geq0}$
the marginal chain, which move according to the kernels $\tilde{P}_{N}$
and $P$, respectively. Let $c_{x}:=1-\|\tilde{P}_{N}(x,\cdot)-P(x,\cdot)\|_{TV}$,
using notions of maximal coupling for random variables defined on a Polish space (see \citealt{Lindvall_2002} and \citealt{Thorisson_2013}). In particular,
there exists a probability measure $\nu_{x}(\cdot)$ such that 
\begin{align*}
P(x,\cdot)\geq c_{x}\nu_{x}(\cdot)\quad\text{and}\quad  \tilde{P}_{N}(x,\cdot)\geq c_{x}\nu_{x}(\cdot).
\end{align*}
Let $c:=\inf_{x\in\mathcal{X}}c_{x}$, define a coupling in the following
way
\begin{itemize}
\item If $\tilde{\Phi}_{n-1}^{N}=\Phi_{n-1}=y$, with probability $c$ draw $\Phi_{n}\sim\nu_{y}(\cdot)$
and set $\tilde{\Phi}_{n}^{N}=\Phi_{n}$. Otherwise, draw independently
$\Phi_{n}\sim R(y,\cdot)$ and $\tilde{\Phi}_{n}^{N}\sim\tilde{R}_{N}(y,\cdot)$,
where
\begin{align*}
& R(y,\cdot):= \left(1-c\right)^{-1}\left(P(y,\cdot)-c\nu_{y}(\cdot)\right)\quad\text{and}\\
& \tilde{R}_{N}(y,\cdot):=  \left(1-c\right)^{-1}\left(\tilde{P}_{N}(y,\cdot)-c\nu_{y}(\cdot)\right).
\end{align*}

\item If $\tilde{\Phi}_{n-1}^{N}\neq \Phi_{n-1}$, draw independently $\Phi_{n}\sim P(y,\cdot)$
and $\tilde{\Phi}_{n}^{N}\sim\tilde{P}_{N}(y,\cdot)$.
\end{itemize}
Since 
\begin{align*}
\mathbb{P}&\left[\tilde{\Phi}_{n}^{N} \neq \Phi_{n}|\tilde{\Phi}_{0}^{N}=\Phi_{0}=x\right]\\
&\leq \mathbb{P}\left[\tilde{\Phi}_{n}^{N}\neq \Phi_{n}|\tilde{\Phi}_{n-1}^{N}=\Phi_{n-1},\tilde{\Phi}_{0}^{N}=\Phi_{0}=x\right] +\mathbb{P}\left[\tilde{\Phi}_{n-1}^{N}\neq \Phi_{n-1}|\tilde{\Phi}_{0}^{N}=\Phi_{0}=x\right]\\
&\leq 1-c+\mathbb{P}\left[\tilde{\Phi}_{n-1}^{N}\neq \Phi_{n-1}|\tilde{\Phi}_{0}^{N}=\Phi_{0}=x\right],
\end{align*}
and noting 
\begin{align*}
\mathbb{P}\left[\tilde{\Phi}_{1}^{N}\neq \Phi_{1}|\tilde{\Phi}_{0}^{N}=\Phi_{0}=x\right] &\leq \sup_{x\in\mathcal{X}}\|\tilde{P}_{N}(x,\cdot)-P(x,\cdot)\|_{TV}\\
&= 1-c,
\end{align*}
an induction argument can be applied to obtain
\begin{align*}
\mathbb{P}\Big[ \tilde{\Phi}_{n}^{N}\neq \Phi_{n}|\tilde{\Phi}_{0}^{N}=\Phi_{0}=x\Big] &\leq n\sup_{x\in\mathcal{X}}\|\tilde{P}_{N}(x,\cdot)-P(x,\cdot)\|_{TV}.
\end{align*}
Therefore, using the coupling inequality, the third term in \eqref{eq:tri.ineq.invariants}
can be bounded by 
\begin{align}
\|\tilde{P}_{N}^{n}(x,\cdot) -P^{n}(x,\cdot)\|_{TV} &\leq \mathbb{P}\left[\tilde{\Phi}_{n}^{N}\neq \Phi_{n}|\tilde{\Phi}_{0}^{N}=\Phi_{0}=x\right] \nonumber \\
&\leq n\sup_{x\in\mathcal{X}}\|\tilde{P}_{N}(x,\cdot)-P(x,\cdot)\|_{TV}.\label{eq:tv.bound.coup}
\end{align}

On the other hand, using the simultaneous geometric ergodicity of
the kernels and provided $N$ is large enough, the noisy and marginal
kernels will each satisfy a geometric drift condition as in \eqref{eq:geom.drift.cond}
with a common drift function $V\geq1$, small set $C$ and constants
$\lambda,b$. Therefore, by \Cref{thm:GE.equiv},
there exist $R>0$, and $\tau<1$ such that 
\begin{align}\label{eq:ge.both.chains}
\|\tilde{P}_{N}^{n}(x,\cdot)-\tilde{\pi}_{N}(\cdot)\|_{TV}\leq RV(x)\tau^{n}\quad\text{and}\quad \|P^{n}(x,\cdot)-\pi(\cdot)\|_{TV}\leq RV(x)\tau^{n}.
\end{align}
Explicit values for $R$ and $\tau$ are in principle possible, as done in \citet{Rosenthal_1995} and \citet{MeynNTweedie_1994}.
For simplicity assume $\inf_{x\in\mathcal{X}}V(x)=1$, then combining
\eqref{eq:tv.bound.coup} and \eqref{eq:ge.both.chains} in \eqref{eq:tri.ineq.invariants},
for all $n\in\mathbb{N}^{+}$ 
\begin{align}
\|\tilde{\pi}_{N}(\cdot) -\pi(\cdot)\|_{TV} &\leq 2R\tau^{n}+n\sup_{x\in\mathcal{X}}\|\tilde{P}_{N}(x,\cdot)-P(x,\cdot)\|_{TV}.
\label{eq:bound.invariants}
\end{align}
So, if an analytic expression in terms of $N$ is available for the second term
on the right hand side of \eqref{eq:bound.invariants}, it will be possible to obtain an 
explicit rate of convergence for $\tilde{\pi}_N$ and $\pi$.

\begin{theorem}
\label{thm:rate.conv.invar}Assume \ref{assm:geom.ergo.marginal},
\ref{assm:minor.condn.subkernel}, \ref{assm:sup.WLLN}
and \ref{assm:sup.inv.moment}. Alternatively, assume \ref{assm:qV.less.KV},
\ref{assm:minor.condn.subkernel} and \ref{assm:sup.WLLN}.
In addition, suppose
\begin{align*}
\sup_{x\in\mathcal{X}}\|\tilde{P}_{N}(x,\cdot)-P(x,\cdot)\|_{TV} \leq \frac{1}{r(N)},
\end{align*}
where $r:\mathbb{N}^{+}\rightarrow\mathbb{R}^{+}$ and $\lim_{N\rightarrow\infty}r(N)=+\infty$.
Then, there exists $D>0$ and $N_{0}\in\mathbb{N}^{+}$ such that
for all $N\geq N_{0}$,
\begin{align*}
\|\tilde{\pi}_{N}(\cdot)-\pi(\cdot)\|_{TV} \leq D\frac{\log\left(r(N)\right)}{r(N)}.
\end{align*}
\end{theorem}

\begin{prooflabel}[]
Let $R>0,\tau\in(0,1)$ and $r>0$. Pick $r$ large enough, such
that
\begin{align*}
\log\left(2Rr\log\left(\tau^{-1}\right)\right) \geq 1,
\end{align*}
then the convex function $f:[1,\infty)\rightarrow\mathbb{R}^{+}$
where
\begin{align*}
f(s) = 2R\tau^{s}+\frac{s}{r},
\end{align*}
is minimised at 
\begin{align*}
s_{*} = \frac{\log\left(2Rr\log\left(\tau^{-1}\right)\right)}{\log\left(\tau^{-1}\right)}.
\end{align*}
Restricting the domain of $f$ to the positive integers and due to
convexity, it is then minimised at either 
\begin{align*}
n_{1} = \lfloor s_{*}\rfloor\qquad\text{or}\qquad n_{2}=\lceil s_{*}\rceil.
\end{align*}
In any case 
\begin{align*}
\min\left\{ f(n_{1}),f(n_{2})\right\} &\leq f(s_{*}+1),\\
&= \frac{1}{r}\left(1+\frac{\tau+\log\left(2Rr\log\left(\tau^{-1}\right)\right)}{\log\left(\tau^{-1}\right)}\right).
\end{align*}

Finally take $N$ large enough such that 
\begin{align*}
\log\left(2Rr(N)\log\left(\tau^{-1}\right)\right)\geq & 1,
\end{align*}
and from \eqref{eq:bound.invariants}
\begin{align*}
\|\tilde{\pi}_{N}(\cdot) -\pi(\cdot)\|_{TV} &\leq \min\left\{ f(n_{1}),f(n_{2})\right\} \\
&\leq \frac{1}{r(N)}\left(1+\frac{\tau+\log\left(2Rr(N)\log\left(\tau^{-1}\right)\right)}{\log\left(\tau^{-1}\right)}\right)\\
&= O\left(\frac{\log\left(r(N)\right)}{r(N)}\right),
\end{align*}
obtaining the result.
\end{prooflabel}

Moreover, when the weights are expressed in terms of arithmetic averages
as in \eqref{eq:wghts.arithm.avg}, an explicit expression for $r(N)$
can be obtained whenever there exists a uniformly bounded moment. This is a slightly stronger assumption than \ref{assm:unif.integ}.
\begin{verse}
\textlabel{(W5)}\label{assm:unif.bdd.moment} There exists $k>0$, such
that the weights $\left\{ W_{x}\right\} _{x}$ satisfy 
\begin{align*}
\sup_{x\in\mathcal{X}}\mathbb{E}_{Q_{x}}\left[W_{x}^{1+k}\right] < \infty.
\end{align*}
\end{verse}

\begin{proposition}
\label{prop:rate.conv.invar}Assume \ref{assm:geom.ergo.marginal},
\ref{assm:minor.condn.subkernel}, \ref{assm:unif.bdd.densities}
and \ref{assm:unif.bdd.moment}. Alternatively, assume \ref{assm:qV.less.KV},
\ref{assm:minor.condn.subkernel} and \ref{assm:unif.bdd.moment}.
Then, there exists $D_{k}>0$ and $N_{0}\in\mathbb{N}^{+}$ such that
for all $N\geq N_{0}$,
\begin{align*}
\|\tilde{\pi}_{N}(\cdot)-\pi(\cdot)\|_{TV} \leq D_{k}\frac{\log\left(N\right)}{N^{1-\frac{2}{2+k}}}.
\end{align*}
If in addition \ref{assm:unif.bdd.moment} holds for all $k>0$, then for any $\varepsilon\in(0,1)$ there will exist $D_{\varepsilon}>0$
and $N_{0}\in\mathbb{N}^{+}$ such that for all $N\geq N_{0}$,
\begin{align*}
\|\tilde{\pi}_{N}(\cdot)-\pi(\cdot)\|_{TV} \leq D_{\varepsilon}\frac{\log\left(N\right)}{N^{1-\varepsilon}}.
\end{align*}
\end{proposition}

\section{Discussion}

In this article, fundamental stability properties of the noisy algorithm have been explored. The noisy Markov kernels considered are perturbed Metropolis--Hastings kernels defined by a collection of state-dependent distributions for non-negative weights all with expectation $1$. The general results do not assume a specific form for these weights, which can be simple arithmetic averages or more complex random variables. The former may arise when unbiased importance sampling estimates of a target density are used, while the latter may arise when such densities are estimated unbiasedly using a particle filter.

Two different sets of sufficient conditions were provided under which the noisy chain inherits geometric ergodicity from the marginal chain. The first pair of conditions, \ref{assm:sup.WLLN} and \ref{assm:sup.inv.moment}, involve a stronger version of the Law of Large Numbers for the weights and uniform convergence of the first negative moment, respectively. For the second set, \ref{assm:sup.WLLN} is still required but \ref{assm:sup.inv.moment} can be replaced with \ref{assm:qV.less.KV}, which imposes a condition on the proposal distribution. These conditions also imply simultaneous geometric ergodicity of a sequence of noisy Markov kernels together with the marginal Markov kernel, which then ensures that the noisy invariant $\tilde{\pi}_{N}$ converges to $\pi$ in total variation as $N$ increases. Moreover, an explicit bound for the rate of convergence between $\tilde{\pi}_{N}$ and $\pi$ is possible whenever an explicit bound (that is uniform in $x$) is available for the convergence between $\tilde{P}_{N}(x,\cdot)$ and $P(x,\cdot)$.

When weights are arithmetic averages as in \eqref{eq:wghts.arithm.avg}, specific conditions were given for inheriting geometric ergodicity from the corresponding marginal chain. The uniform integrability condition in \ref{assm:unif.integ} ensures that \ref{assm:sup.WLLN} is satisfied, whereas \ref{assm:unif.bdd.densities} is essential for satisfying \ref{assm:sup.inv.moment}. Regarding the noisy invariant distribution $\tilde{\pi}_{N}$, \ref{assm:unif.bdd.moment}, which is slightly stronger than \ref{assm:unif.integ}, leads to an explicit bound on the rate of convergence of this distribution to $\pi$.

The noisy algorithm remains undefined when the weights have positive probability of being zero. If both weights were zero one could accept the move, reject the move or keep sampling new weights until one of them is not zero. Each of these lead to different behaviour.

As seen in the examples of \Cref{sub:Remarks.results}, the behaviour of the ratio of the weights (at least in the tails of the target) plays an important role in the ergodic properties of the noisy chain. In this context, it seems plausible to obtain geometric noisy chains, even when the marginal is not, if the ratio of the weights decays sufficiently fast to zero in the tails. Another interesting possibility, that may lead to future research, is to relax the condition on the expectation of the weights to be identically one.

\section*{Acknowledgements}
All authors would like to thank the EPSRC-funded Centre for Research in Statistical methodology (EP/D002060/1). The first author was also supported by Consejo Nacional de Ciencia y Tecnolog\'ia. The third author's research was also supported by the EPSRC Programme Grant {\em ilike} (EP/K014463/1).

\bibliographystyle{abbrvnat}
\bibliography{Biblio}

\appendix
\section{Proofs}\label{sec:Proofs}

\subsection{On state-dependent random walks}

The following proposition for state-dependent Markov chains on the positive
integers will be useful for addressing some proofs. See \citet{Norris_1999}
for a proof of parts $(i)$ and $(ii)$, for part $(iii)$ see \citet{Callaert_1973}, which is proved within the birth-death process context.
\begin{proposition}
\label{prop:non.hom.rw}Suppose we have a random walk $\Phi$ on $\mathbb{N}^{+}$
with transition kernel $P$. Define for $m\geq1$ 
\begin{align*}
p_{m} := P(m,\left\{ m+1\right\} )\quad\text{and}\quad q_{m}:=P(m,\left\{ m-1\right\} ),
\end{align*}
with $q_{1}=0,p_{1}\in(0,1]$ and $p_{m},q_{m}>0,p_{m}+q_{m}\leq1$
for all $m\geq2$. The resulting chain is:\renewcommand{\labelenumi}{\roman{enumi}.}
\begin{enumerate}
\item recurrent if and only if
\begin{align*}
\sum_{m=2}^{\infty}\prod_{i=2}^{m}\frac{q_{i}}{p_{i}}\rightarrow  \infty;
\end{align*}

\item positive recurrent if and only if 
\begin{align*}
\sum_{m=2}^{\infty}\prod_{i=2}^{m}\frac{p_{i-1}}{q_{i}} < \infty;
\end{align*}

\item geometrically ergodic if
\begin{align*}
\lim_{m\rightarrow\infty}p_{m} < \lim_{m\rightarrow\infty}q_{m}.
\end{align*}

\end{enumerate}
\end{proposition}
\begin{remark}
Notice that (3) is not an if and only if statement and that it implies
(2). Additionally, if the chain is not state-dependent, (2) implies (3).
\end{remark}

\subsection{\Cref{sec:Examples}}
\begin{prooflabel}[\Cref{prop:ex.logconc.hom}]
Since $h$ is convex
\begin{align*}
h(m)-h(m-1) \geq h'(m-1) \quad \text{and} \quad h(m)-h(m+1) \leq -h'(m),
\end{align*}
implying
\begin{align*}
\frac{\tilde{P}_{N}(m,\left\{ m-1\right\} )}{\tilde{P}_{N}(m,\left\{ m+1\right\} )} &\geq \frac{\mathbb{E}\left[\min\left\{ 1,\exp\{h'(m-1)\}\frac{W_{N}^{(1)}}{W_{N}^{(2)}}\right\} \right]}{\mathbb{E}\left[\min\left\{ 1,\exp\{-h'(m)\}\frac{W_{N}^{(1)}}{W_{N}^{(2)}}\right\} \right]}.
\end{align*}
Define $Z:=\frac{W_{N}^{(1)}}{W_{N}^{(2)}}$, and since $\pi(m)\rightarrow0$
it is true that 
\begin{align}\label{eq:logk.positive}
\log(k):=\lim_{m\rightarrow\infty}h'(m) > 0,
\end{align}
hence
\begin{align}
\lim_{m\rightarrow\infty}\frac{\tilde{P}_{N}(m,\left\{ m-1\right\} )}{\tilde{P}_{N}(m,\left\{ m+1\right\} )} \geq \frac{\mathbb{E}\left[\min\left\{ 1,kZ\right\} \right]}{\mathbb{E}\left[\min\left\{ 1,k^{-1}Z\right\} \right]}.\label{eq:ex1.ratio.probs}
\end{align}

If $k=+\infty$, it is clear that the limit in \eqref{eq:ex1.ratio.probs} diverges,
consequently the noisy chain is geometrically ergodic according to
\Cref{prop:non.hom.rw}. If $k<\infty$, the noisy chain
will be geometrically ergodic if
\begin{align*}
\mathbb{E}\left[\min\left\{ 1,kZ\right\} \right] > \mathbb{E}\left[\min\left\{ 1,k^{-1}Z\right\} \right],
\end{align*}
which can be translated to
\begin{align*}
k\mathbb{E}\left[Z\mathds{1}_{\{Z\leq k^{-1}\}}\right]+&\mathbb{P}\left[Z>k^{-1}\right] &>k^{-1}\mathbb{E}\left[Z\mathds{1}_{\{Z<k\}}\right]+\mathbb{P}\left[Z\geq k\right],
\end{align*}
or equivalently to 
\begin{align}
k\mathbb{P}\left[k^{-1}<Z<k\right]+\left(k^{2}-1\right) \mathbb{E}\left[Z\mathds{1}_{\left\{ Z\leq k^{-1}\right\} }\right] > \mathbb{E}\left[Z\mathds{1}_{\left\{ k^{-1}<Z<k\right\} }\right].
\label{eq:ex1.condn}
\end{align}
Now consider two cases, first if $\mathbb{P}\left[k^{-1}<Z<k\right]>0$
then it is clear that
\begin{align*}
\mathbb{E}\left[\left(k-Z\right)\mathds{1}_{\left\{ k^{-1}<Z<k\right\} }\right] > 0,
\end{align*}
which satisfies \eqref{eq:ex1.condn}. Finally, if $\mathbb{P}\left[k^{-1}<Z<k\right]=0$
then
\begin{align*}
\mathbb{P}\left[Z\leq k^{-1}\right] = \frac{1}{2}=\mathbb{P}\left[Z\geq k\right],
\end{align*}
implying from \eqref{eq:logk.positive}
\begin{align*}
\left(k^{2}-1\right)\mathbb{E}\left[Z\mathds{1}_{\left\{ Z\leq k^{-1}\right\} }\right] > 0,
\end{align*}
and leading to \eqref{eq:ex1.condn}.
\end{prooflabel}

\begin{prooflabel}[\Cref{prop:trans.geom.hom}]
For simplicity the subscript $N$ is dropped. In this case, 
\begin{align*}
Q_{m,1} =Q=\mathcal{L}\left((b-\varepsilon)Ber(s)+\varepsilon\right),
\end{align*}
and the condition $\mathbb{E}_{Q}\left[W\right]=1$ implies 
\begin{align}
s= \frac{1-\varepsilon}{b-\varepsilon}.\label{eq:ex1.s}
\end{align}
Let $\theta\in\left(\frac{1}{1+2\varepsilon},1\right)$ and set 
\begin{align}
b= \varepsilon\frac{2\theta}{1-\theta},\label{eq:ex1.b}
\end{align}
this implies $\bar{\alpha}(m,w;m-1,u)\equiv1$ and 
\begin{align*}
\bar{\alpha}(m,w;m+1,u) &= \begin{cases}
\begin{array}{c}
\frac{1-\theta}{2\theta}\\
1\\
\left(\frac{1-\theta}{2\theta}\right)^{2}
\end{array}\quad\text{if} & \begin{array}{c}
u=w\\
u=b,w=\varepsilon\\
u=\varepsilon,w=b
\end{array}\end{cases}.
\end{align*}
Therefore, for $m\geq2$,
\begin{align*}
\tilde{\alpha}(m, m-1) &= 1\quad\text{and}\\
\tilde{\alpha}(m, m+1) &= \frac{1-\theta}{2\theta}\left(s^{2}+\left(1-s\right)^{2}\right) +\left(1+\left(\frac{1-\theta}{2\theta}\right)^{2}\right)s\left(1-s\right).
\end{align*}
Consequently, $\tilde{P}(m,\left\{ m-1\right\} )=1-\theta$ and 
\begin{align*}
\tilde{P}(m, \left\{ m+1\right\} ) &= \theta\left(\frac{1-\theta}{2\theta}\left(s^{2}+\left(1-s\right)^{2}\right)+\left(1+\left(\frac{1-\theta}{2\theta}\right)^{2}\right)s\left(1-s\right)\right)\\
&> \theta s(1-s).
\end{align*}

From \Cref{prop:non.hom.rw}, if 
\begin{align*}
\tilde{P}(m,\left\{ m+1\right\} ) > \tilde{P}(m,\left\{ m-1\right\}),
\end{align*}
then the noisy chain will be transient. For this to happen, it is
enough to pick $\theta$ and $s$ such that 
\begin{align*}
\theta s(1-s)-(1-\theta)\geq 0.
\end{align*}
Let $s=\varepsilon$, then from \eqref{eq:ex1.s} and \eqref{eq:ex1.b}
\begin{align}
\theta &= \frac{(1-\varepsilon+\varepsilon^{2})}{1-\varepsilon+3\varepsilon^{2}}\label{eq:ex1.q}\\
&= 1-\frac{2\varepsilon^{2}}{1-\varepsilon+3\varepsilon^{2}},\nonumber 
\end{align}
 and if $\varepsilon\leq2-\sqrt{3}$ then
\begin{align*}
\theta s(1-s)-(1-\theta) &= \frac{\varepsilon}{1-\varepsilon+3\varepsilon^{2}}\left((1-\varepsilon+\varepsilon^{2})(1-\varepsilon)-2\varepsilon\right)\\
&\geq \frac{\varepsilon}{1-\varepsilon+3\varepsilon^{2}}\left((1-\varepsilon)^{2}-2\varepsilon\right)\\
&= \frac{\varepsilon}{1-\varepsilon+3\varepsilon^{2}}\left((2-\varepsilon)^{2}-3\right)\\
&\geq 0.
\end{align*}

Hence, for $\varepsilon\in\left(0,2-\sqrt{3}\right)$ and setting
$s=\varepsilon$, $\theta$ as in \eqref{eq:ex1.q} and $b$ as
in \eqref{eq:ex1.b}, the resulting noisy chain is transient.
\end{prooflabel}

\begin{prooflabel}[\Cref{prop:trans.geom.inhom}]
For simplicity the subscript $N$ is dropped. In this case, 
\begin{align*}
Q_{m,1} = Q_{m}=\mathcal{L}\left((b-\varepsilon_{m})Ber(s_{m})+\varepsilon_{m}\right),
\end{align*}
and the condition $\mathbb{E}_{Q_{m}}\left[W_{m}\right]=1$ implies
\begin{align*}
s_{m} = \frac{1-\varepsilon_{m}}{b-\varepsilon_{m}}.
\end{align*}
Then, for $m$ large enough 
\begin{align*}
\tilde{\alpha}(m,m-1) &= \mathbb{E}\left[\bar{\alpha}(m,W_{m};m-1,W_{m-1})\right]\\
&= \min\left\{ 1,\frac{2\theta}{1-\theta}\right\} s_{m-1}s_{m}+s_{m-1}(1-s_{m}) +(1-s_{m-1})(1-s_{m})\mathds{1}_{\left\lbrace m\Mod{3}=0\right\rbrace }\\
 &\quad +O\left(m^{-1}\right)\quad\text{and}\\
\tilde{\alpha}(m,m-1) &= \mathbb{E}\left[\bar{\alpha}(m,W_{m};m-1,W_{m-1})\right]\\
&= \min\left\{ 1,\frac{1-\theta}{2\theta}\right\} s_{m}s_{m+1}+(1-s_{m})s_{m+1} +(1-s_{m})(1-s_{m+1})\mathds{1}_{\left\lbrace m\Mod{3}\neq2\right\rbrace }\\
 &\quad +O\left(m^{-1}\right).
\end{align*}

Define 
\begin{align*}
c_{m} &:= \frac{\tilde{P}(m,\left\{ m-1\right\} )}{\tilde{P}(m,\left\{ m+1\right\} )}\\
&= \frac{(1-\theta)\tilde{\alpha}(m,m-1)}{\theta\tilde{\alpha}(m,m+1)}.
\end{align*}
 Since $s_{m}\rightarrow\frac{1}{b}$ as $m\rightarrow\infty$, 
\begin{align*}
c_{0,\infty} &:= \lim_{k\rightarrow\infty}c_{3k}\\
&= \left(\frac{1-\theta}{\theta}\right) \frac{\left(\min\left\{ 1,\frac{2\theta}{1-\theta}\right\} -1\right)\frac{1}{b^{2}}+\frac{1}{b}+\left(1-\frac{1}{b}\right)^{2}}{\left(\min\left\{ 1,\frac{1-\theta}{2\theta}\right\} -1\right)\frac{1}{b^{2}}+\frac{1}{b}+\left(1-\frac{1}{b}\right)^{2}}\\
&\leq \left(\frac{1-\theta}{\theta}\right)\frac{1}{1-\frac{1}{b}}\\
&= \left(\frac{1-\theta}{\theta}\right)\frac{b}{b-1}=:l_{0},
\end{align*}
\begin{align*}
c_{1,\infty}:= & \lim_{k\rightarrow\infty}c_{3k}c_{3k+1}\\
&= c_{0,\infty}\left(\frac{1-\theta}{\theta}\right) \frac{\left(\min\left\{ 1,\frac{2\theta}{1-\theta}\right\} -1\right)\frac{1}{b^{2}}+\frac{1}{b}}{\left(\min\left\{ 1,\frac{1-\theta}{2\theta}\right\} -1\right)\frac{1}{b^{2}}+\frac{1}{b}+\left(1-\frac{1}{b}\right)^{2}}\\
&\leq l_{0}\left(\frac{1-\theta}{\theta}\right)\frac{\frac{1}{b}}{1-\frac{1}{b}}\\
&= \left(\frac{1-\theta}{\theta}\right)^{2}\frac{b}{(b-1)^{2}}=:l_{1}
\end{align*}
and
\begin{align*}
\lim_{k\rightarrow\infty} c_{3k}c_{3k+1}c_{3k+2} &= c_{1,\infty}\left(\frac{1-\theta}{\theta}\right)\frac{\left(\min\left\{ 1,\frac{2\theta}{1-\theta}\right\} -1\right)\frac{1}{b^{2}}+\frac{1}{b}}{\left(\min\left\{ 1,\frac{1-\theta}{2\theta}\right\} -1\right)\frac{1}{b^{2}}+\frac{1}{b}}\\
&\leq l_{1}\left(\frac{1-\theta}{\theta}\right)\frac{\frac{1}{b}}{\frac{1}{b}\left(1-\frac{1}{b}\right)}\\
&= \left(\frac{1-\theta}{\theta}\right)^{3}\frac{b^{2}}{\left(b-1\right)^{3}}=:l_{2}.
\end{align*}

Therefore, for any $\delta>0$ there exists $k_{0}\in\mathbb{N}^{+}$,
such that whenever $k\geq k_{0}+1$
\begin{align*}
K &:= \prod_{j=k_{0}}^{k-1}c_{3j}c_{3j+1}c_{3j+2}\\
&< (l_{2}+\delta)^{k-k_{0}},
\end{align*}
implying
\begin{align*}
 & Kc_{3k}<(l_{2}+\delta)^{k-k_{0}}(l_{0}+\delta),\\
 & Kc_{3k}c_{3k+1}<(l_{2}+\delta)^{k-k_{0}}(l_{1}+\delta)\quad\text{and}\\
 & Kc_{3k}c_{3k+1}c_{3k+2}<(l_{2}+\delta)^{k-k_{0}}(l_{2}+\delta).
\end{align*}
Hence, for $i\in\lbrace0,1,2\rbrace$ and some $C>0$ 
\begin{align*}
\prod_{j=2}^{3k+i}c_{j} &< C(l_{2}+\delta)^{k}.
\end{align*}

Let $a_{m}:=\prod_{j=2}^{m}c_{j}$, then a sufficient condition for
the series $\sum_{m=2}^{\infty}a_{m}$ to converge, implying a transient
chain according to \Cref{prop:non.hom.rw}, is $l_{2}<1$.
This is the case for $b\geq3+\left(\frac{1-\theta}{\theta}\right)^{3}$,
since 
\begin{align*}
1-l_{2} &= 1-\left(\frac{1-\theta}{\theta}\right)^{3}\frac{b^{2}}{\left(b-1\right)^{3}}\\
&= \frac{b^{2}}{\left(b-1\right)^{3}}\left(\frac{\left(b-1\right)^{3}}{b^{2}}-\left(\frac{1-\theta}{\theta}\right)^{3}\right)\\
&= \frac{b^{2}}{\left(b-1\right)^{3}}\left(b-3+\frac{3}{b}-\frac{1}{b^{2}}-\left(\frac{1-\theta}{\theta}\right)^{3}\right)\\
&> \frac{b^{2}}{\left(b-1\right)^{3}}\left(b-3-\left(\frac{1-\theta}{\theta}\right)^{3}\right)\\
&\geq 0.
\end{align*}
Hence, the resulting noisy chain is transient if $b\geq3+\left(\frac{1-\theta}{\theta}\right)^{3}$,
for any $\theta\in(0,1)$.
\end{prooflabel}

\subsection{\Cref{sec:Inheritance}}
\begin{prooflabel}[\Cref{lem:sup.probs}]
For any $\delta >0$
\begin{align*}
\mathbb{P}\left[\frac{W_{z,N}}{W_{x,N}}\leq 1-\delta\right] &\leq \mathbb{P}\left[W_{x,N}\geq 1+\frac{\delta}{2}\right]+ \mathbb{P}\left[W_{z,N}\leq 1-\frac{\delta}{2}\right]\\
&\leq \mathbb{P}\left[\Big\vert W_{x,N}-1 \Big\vert \geq \frac{\delta}{2}\right]+ \mathbb{P}\left[\Big\vert W_{z,N}-1 \Big\vert \geq \frac{\delta}{2}\right]\\
&\leq 2\sup_{x\in\mathcal{X}}\mathbb{P}_{Q_{x,N}}\left[\Big\vert W_{x,N}-1\Big\vert\geq\frac{\delta}{2}\right].
\end{align*}
\end{prooflabel}

\begin{prooflabel}[\Cref{lem:rejec.probs}]
Using the inequality 
\begin{align*}
\min\left\{ 1,ab\right\} \geq\min\left\{ 1,a\right\} \min\left\{ 1,b\right\}, \quad \text{for } a,b\geq0,
\end{align*}
and applying Markov's inequality with $\delta>0$,
\begin{align*}
\tilde{\rho}_{N}(x) &= 1- \int_{\mathcal{X}} q(x,dz) \tilde{\alpha}_N(x,z) \\
&\leq 1-\int_{\mathcal{X}}q(x,dz)\alpha(x,z)\mathbb{E}\left[\min\left\{ 1,\frac{W_{z,N}}{W_{x,N}}\right\} \right]\\
&\leq 1-\left(1-\delta\right)\int_{\mathcal{X}}q(x,dz)\alpha(x,z)\mathbb{P}\left[\min\left\{ 1,\frac{W_{z,N}}{W_{x,N}}\right\} >1-\delta\right]\\
&= 1-\left(1-\delta\right)\int_{\mathcal{X}}q(x,dz)\alpha(x,z)+\left(1-\delta\right) \int_{\mathcal{X}}q(x,dz)\alpha(x,z)\mathbb{P}\left[ \frac{W_{z,N}}{W_{x,N}}\leq 1-\delta\right]\\
&\leq 1-\left(1-\delta\right)\left(1-\rho(x)\right) +\int_{\mathcal{X}}q(x,dz)\alpha(x,z)\mathbb{P}\left[\frac{W_{z,N}}{W_{x,N}}\leq1-\delta\right].
\end{align*}
Finally, using \Cref{lem:sup.probs}
\begin{align*}
\tilde{\rho}_{N}(x) &\leq \rho(x)+\delta\left(1-\rho(x)\right) +2\sup_{x\in\mathcal{X}}\mathbb{P}\left[\big\vert W_{x,N}-1\big\vert\geq\frac{\delta}{2}\right]\left(1-\rho(x)\right)\\
&\leq \rho(x)+\delta+2\sup_{x\in\mathcal{X}}\mathbb{P}\left[\big\vert W_{x,N}-1\big\vert\geq\frac{\delta}{2}\right].
\end{align*}
\end{prooflabel}

\begin{prooflabel}[\Cref{lem:accep.probs.prop}]
For the first claim apply Jensen's inequality and the fact that 
\begin{align*}
\min\left\{ 1,ab\right\} \leq\min\left\{ 1,a\right\} b, \quad \text{for  } a\geq0 \text{ and } b\geq1,
\end{align*}
hence
\begin{align*}
\tilde{\alpha}_{N}(x,z) &\leq \min\left\{1,\frac{\pi(z)q(z,x)}{\pi(x)q(x,z)}\mathbb{E}\left[\frac{W_{z,N}}{W_{x,N}}\right]\right\}\\
&\leq \alpha(x,z)\mathbb{E}\left[W_{x,N}^{-1}\right]\mathbb{E}\left[W_{z,N}\right].
\end{align*}
\end{prooflabel}

\begin{prooflabel}[\Cref{lem:accept.probs.linear}]
Using the inequality 
\begin{align*}
\min\left\{ 1,ab\right\} \leq\min\left\{ 1,a\right\} b, \quad \text{for  } a\geq0 \text{ and } b\geq1,
\end{align*}
\begin{align*}
\tilde{\alpha}_{N} (x,z) &= \mathbb{E}\left[\bar{\alpha}_{N}(x,W_{x,N};z,W_{z,N})\mathds{1}_{\left\{ \frac{W_{z,N}}{W_{x,N}}<1+\eta\right\} }\right] +\mathbb{E}\left[\bar{\alpha}_{N}(x,W_{x,N};z,W_{z,N})\mathds{1}_{\left\{ \frac{W_{z,N}}{W_{x,N}}\geq1+\eta\right\} }\right]\\
&\leq \alpha(x,z)\left(1+\eta\right)\mathbb{P}\left[\frac{W_{z,N}}{W_{x,N}}<1+\eta\right] +\mathbb{P}\left[\frac{W_{z,N}}{W_{x,N}}\geq1+\eta\right]\\
&\leq \alpha(x,z)+\eta+\mathbb{P}\left[\frac{W_{z,N}}{W_{x,N}}\geq1+\eta\right],
\end{align*}
Notice that
\begin{align*}
\mathbb{P}\left[\frac{W_{z,N}}{W_{x,N}}\geq1+\eta\right]&=\mathbb{P}\left[\frac{W_{x,N}}{W_{z,N}}\leq \frac{1}{1+\eta} \right],
\end{align*}
then applying \Cref{lem:sup.probs} taking $\delta=\frac{\eta}{1+\eta}$.
\begin{align*}
\tilde{\alpha}_{N} & (x,z) \leq  \alpha(x,z)+\eta+2\sup_{x\in\mathcal{X}}\mathbb{P}\left[\Big\vert W_{x,N}-1\Big\vert\geq\frac{\eta}{2\left(1+\eta\right)}\right].
\end{align*}
\end{prooflabel}

\begin{prooflabel}[\Cref{prop:log.Lipschitz}]
Taking $V=\pi^{-s}$, where $0<s<\min\left\{ 1,\frac{a}{L}\right\} $,
\begin{align*}
\frac{qV(x)}{V(x)} &= \int_{\mathcal{X}}\frac{V(z)}{V(x)}q(x,dz)\\
&= \int_{\mathcal{X}}\left(\frac{\pi(x)}{\pi(z)}\right)^{s}q(x,dz)\\
&\leq \int_{\mathbb{R}^{d}}\exp\left\{ a\|z-x\|\right\} q(\|z-x\|)dz.
\end{align*}
Finally, using the transformation $u=z-x$,
\begin{align*}
\frac{qV(x)}{V(x)} &\leq \int_{\mathbb{R}^{d}}\exp\left\{ a\|u\|\right\} q(\|u\|)du,
\end{align*}
which implies \ref{assm:qV.less.KV}.
\end{prooflabel}

\begin{prooflabel}[\Cref{prop:convergence.fatous}]
By properties of the arithmetic and harmonic means
\begin{align*}
\frac{1}{N}\sum_{k=1}^{N}\frac{1}{W_{x}^{(k)}}-\frac{N}{\sum_{k=1}^{N}W_{x}^{(k)}} \geq 0,
\end{align*}
which implies, by Jensen's inequality,
\begin{align*}
\mathbb{E}\left[\frac{1}{N}\sum_{k=1}^{N}\frac{1}{W_{x}^{(k)}}-\frac{N}{\sum_{k=1}^{N}W_{x}^{(k)}}\right] &\leq \mathbb{E}\left[W_{x}^{-1}\right]-1.
\end{align*}
Then, using Fatou's lemma and the law of large numbers
\begin{align*}
\mathbb{E}\left[W_{x}^{-1}\right]-1 &\geq \limsup_{N\rightarrow\infty}\mathbb{E}\left[\frac{1}{N}\sum_{k=1}^{N}\frac{1}{W_{x}^{(k)}}-W_{x,N}^{-1}\right]\\
&\geq \liminf_{N\rightarrow\infty}\mathbb{E}\left[\frac{1}{N}\sum_{k=1}^{N}\frac{1}{W_{x}^{(k)}}-W_{x,N}^{-1}\right]\\
&\geq \mathbb{E}\left[\liminf_{N\rightarrow\infty}\left(\frac{1}{N}\sum_{k=1}^{N}\frac{1}{W_{x}^{(k)}}\right)-\limsup_{N\rightarrow\infty}W_{x,N}^{-1}\right]\\
&\geq \mathbb{E}\left[W_{x}^{-1}\right]-1,
\end{align*}
hence
\begin{align}
\lim_{N\rightarrow\infty} & \mathbb{E}\left[\frac{1}{N}\sum_{k=1}^{N}\frac{1}{W_{x}^{(k)}}-W_{x,N}^{-1}\right]=E\left[W_{x}^{-1}\right]-1.\label{eq:proof.expec}
\end{align}
Finally, since 
\begin{align*}
\mathbb{E}\left[\frac{1}{N}\sum_{k=1}^{N}\frac{1}{W_{x}^{(k)}}-W_{x,N}^{-1}\right] &= E\left[W_{x}^{-1}\right]-E\left[W_{x,N}^{-1}\right],
\end{align*}
the expression in \eqref{eq:proof.expec} becomes
\begin{align*}
\lim_{N\rightarrow\infty} & \mathbb{E}\left[W_{x,N}^{-1}\right]=1.
\end{align*}
\end{prooflabel}

\begin{prooflabel}[\Cref{lem:sums.rvs}]
The proof of $(i)$ is motivated by \citet[][Theorem~2.1]{PiegorschNCasella_1985} and \citet[][Theorem~3]{KhuriNCasella_2002}, however the existence of a density function is not assumed here. Since $Z^{-p}$ is positive,
\begin{align*}
\mathbb{E}\left[ Z^{-p} \right]&= \int_{\mathbb{R}^+}\mathbb{P}\left[ Z^{-p} \geq z \right]dz\\
&\leq \frac{1}{\gamma^p} + \int_{\left(\gamma^{-p}, \infty \right)}\mathbb{P}\left[ Z^{-p} \geq z \right]dz\\
&= \frac{1}{\gamma^p} + \int_{\left(0, \gamma \right)} pu^{-p-1} \mathbb{P}\left[ Z \leq u \right]du\\
&\leq \frac{1}{\gamma^p} + pM\frac{\gamma^{\alpha-p}}{\alpha-p}.
\end{align*}

For part $(ii)$, since the random variables $\left\{ Z_{i}\right\}$ are positive, then for any $z>0$
\begin{align*}
\mathbb{P}\left[\sum_{i=1}^{N}Z_{i}\leq z\right] & =\mathbb{P}\left[\sum_{i=1}^{N}Z_{i}\leq z,\max_{i\in\{1,\dots,N\}}\left\{ Z_{i}\right\} \leq z\right].
\end{align*}
Therefore, for $z\in(0,\gamma)$
\begin{align*}
\mathbb{P}\left[\sum_{i=1}^{N}Z_{i}\leq z\right] & \leq\mathbb{P}\left[\max_{i\in\{1,\dots,N\}}\left\{ Z_{i}\right\} \leq z\right]\\
 & =\prod_{i=1}^{N}\mathbb{P}\left[Z_{i}\leq z\right]\\
 & \leq\prod_{i=1}^{N}M_{i}z^{\sum_{i=1}^{N}\alpha_{i}}.
\end{align*}

Part (iii) can be seen as a consequence of $W_{x,N}$ and $W_{x,N+1}$ being convex ordered and $g(x)=x^{-p}$ being a convex function for $x>0$ and $p\geq0$, \citep[see, e.g.,][]{AndrieuNVihola_2014}. We provide a self-contained proof by defining for $j\in\{1,\dots,N+1\}$
\begin{align*}
S_{x,N}^{(j)}:=\frac{1}{N} \sum_{k=1,k\neq j}^{N+1} W_{x}^{(k)},
\end{align*}
and we have
\begin{align*}
W_{x,N+1}=\frac{1}{N+1} \sum_{j=1}^{N+1} S_{x,N}^{(j)}
\end{align*}
and since the arithmetic mean is greater than or equal to the geometric mean
\begin{align*}
W_{x,N+1}\geq \left( \prod_{j=1}^{N+1} S_{x,N}^{(j)} \right)^{\frac{1}{N+1}}.
\end{align*}
This implies for $p>0$
\begin{align*}
\mathbb{E}\left[ W_{x,N+1}^{-p} \right] &\leq \mathbb{E}\left[ \left( \prod_{j=1}^{N+1} S_{x,N}^{(j)} \right)^{-\frac{p}{N+1}} \right]\\
&\leq \prod_{j=1}^{N+1} \left( \mathbb{E}\left[ \left( S_{x,N}^{(j)} \right)^{-p} \right] \right)^{\frac{1}{N+1}}\\
&= \mathbb{E}\left[ \left( S_{x,N}^{(1)} \right)^{-p} \right]\\
&= \mathbb{E}\left[ W_{x,N}^{-p} \right],
\end{align*}
where H\"older's inequality has been used and the fact that the random variables $\left\{S_{x,N}^{(j)} : j \in {1,\dots,N+1} \right\}$ are identically distributed according to $Q_{x,N}$.

For part $(iv)$, let $M_{\gamma}=\sup_{y\in[\gamma,\infty)}|g(y)|$ and due to continuity at $y=1$, for
any $\varepsilon>0$ there exists a $\delta>0$ such that
\begin{align*}
\mathbb{E}\left[\big\vert g{\left(W_{x,N}\right)-g(1)\big\vert\mathds{1}_{W_{x,N}\in[\gamma,\infty)}}\right ] &\leq 2M_{\gamma}\mathbb{P}\left[\gamma\leq W_{x,N}\leq 1-\delta\right] +2M_{\gamma}\mathbb{P}\left[1+\delta \leq W_{x,N}\right]\\
 &\quad+\mathbb{E}\left[\big\vert g\left(W_{x,N}\right)-g(1)\big\vert\mathds{1}_{W_{x,N}\in\left(1-\delta,1+\delta\right)}\right]\\
&\leq 2M_{\gamma}\mathbb{P}\left[\big\vert W_{x,N}-1\big\vert\geq\delta\right]+\varepsilon\mathbb{P}\left[\big\vert W_{x,N}-1\big\vert < \delta\right].
\end{align*}
Therefore, for fixed $\varepsilon$ and by \ref{assm:sup.WLLN}
\begin{align*}
\lim_{N\rightarrow\infty}\sup_{x\in\mathcal{X}} \mathbb{E} \left[\big\vert g\left(W_{x,N}\right)-g(1)\big\vert\mathds{1}_{W_{x,N}\in[\gamma,\infty)}\right] &\leq 2M_{\gamma}\lim_{N\rightarrow\infty}\sup_{x\in\mathcal{X}}\mathbb{P}\left[\big\vert W_{x,N}-1\big\vert\geq\delta\right]+\varepsilon\\
&\leq \varepsilon,
\end{align*}
obtaining the result since $\varepsilon$ can be picked arbitrarily
small.
\end{prooflabel}

\begin{prooflabel}[\Cref{prop:GE.geom.inhom.allN}]
First notice that if $l<\infty$ then $l\geq1$. To see this, define
\begin{align*}
a_{m} := \frac{\varepsilon_{m-1}}{\varepsilon_{m}},
\end{align*}
then for fixed $\delta>0$, there exists $M\in\mathbb{N}$ such that
for $m\geq M$ 
\begin{align*}
a_{m} < l+\delta.
\end{align*}
Then, for $m\geq M$ 
\begin{align*}
\frac{\varepsilon_{1}}{\varepsilon_{m}} &= \prod_{j=2}^{m}a_{j}\\
&< \left(l+\delta\right)^{m-M}\frac{\varepsilon_{1}}{\varepsilon_{M}},
\end{align*}
and because $\varepsilon_{m}\rightarrow0$, it is clear that $\left(l+\delta\right)^{m}\rightarrow\infty$
as $m\rightarrow\infty$. Therefore, $l+\delta>1$ and since $\delta$
can be taken arbitrarily small, it is true that $l\geq1$.

Now, for weights as in \eqref{eq:weight.binom} and using a simple
random walk proposal, the noisy acceptance probability can be expressed
as

\begin{align}
\begin{split}\tilde{\alpha}_{N} (m,m-1) &= \sum_{j=0}^{N}\sum_{k=0}^{N}\min\left\{ 1,\frac{2\theta}{1-\theta}\frac{b_{m-1}j+\left(N-j\right)\varepsilon_{m-1}}{b_{m}k+\left(N-k\right)\varepsilon_{m}}\right\} \binom{N}{j}\\
 &\quad \times\binom{N}{k}\left(s_{m-1}\right)^{j}\left(s_{m}\right)^{k}\left(1-s_{m-1}\right)^{N-j}\left(1-s_{m}\right)^{N-k}
\end{split}
\label{eq:ex3.expect1}
\end{align}
and
\begin{align}
\begin{split}\tilde{\alpha}_{N} (m,m+1) &= \sum_{j=0}^{N}\sum_{k=0}^{N}\min\left\{ 1,\frac{1-\theta}{2\theta}\frac{b_{m+1}j+\left(N-j\right)\varepsilon_{m+1}}{b_{m}k+\left(N-k\right)\varepsilon_{m}}\right\} \binom{N}{j}\\
 &\quad \times\binom{N}{k}\left(s_{m+1}\right)^{j}\left(s_{m}\right)^{k}\left(1-s_{m+1}\right)^{N-j}\left(1-s_{m}\right)^{N-k}.
\end{split}
\label{eq:ex3.expect2}
\end{align}

Since $b_{m}\rightarrow\infty$, then $s_{m}\rightarrow0$ as $m\rightarrow\infty$;
therefore, any term in \eqref{eq:ex3.expect1} and \eqref{eq:ex3.expect2},
for which $j+k\neq0$, tends to zero as $m\rightarrow\infty$. Hence,
\begin{align*}
\tilde{\alpha}_{N} (m,m-1) &= \min\left\{ 1,\frac{2\theta}{1-\theta}\times\frac{\varepsilon_{m-1}}{\varepsilon_{m}}\right\} \left(1-s_{m-1}\right)^{N}\left(1-s_{m}\right)^{N}+o(1)
\end{align*}
and 
\begin{align*}
\tilde{\alpha}_{N} (m,m+1) &= \min\left\{ 1,\frac{1-\theta}{2\theta}\times\frac{\varepsilon_{m+1}}{\varepsilon_{m}}\right\} \left(1-s_{m+1}\right)^{N}\left(1-s_{m}\right)^{N}+o(1),
\end{align*}
implying,
\begin{align}
\lim_{m\rightarrow\infty} \frac{\tilde{P}_{N}\left(m,\left\{ m-1\right\} \right)}{\tilde{P}_{N}\left(m,\left\{ m+1\right\} \right)} &= \frac{\left(1-\theta\right)\lim_{m\rightarrow\infty}\min\left\{1,\frac{2\theta}{1-\theta}\times\frac{\varepsilon_{m-1}}{\varepsilon_{m}}\right\}}{\theta\lim_{m\rightarrow\infty}\min\left\{1,\frac{1-\theta}{2\theta}\times\frac{\varepsilon_{m+1}}{\varepsilon_{m}}\right\}}.
\label{eq:ex.ratio.P}
\end{align}
If $l=+\infty$, \eqref{eq:ex.ratio.P} tends to $+\infty$, whereas
if $l<\infty$ 
\begin{align*}
\lim_{m\rightarrow\infty}\frac{\tilde{P}_{N}\left(m,\left\{ m-1\right\} \right)}{\tilde{P}_{N}\left(m,\left\{ m+1\right\} \right)} &= 2l\frac{\min\left\{1-\theta,2\theta l\right\}}{\min\left\{2\theta l,1-\theta\right\}}\\
&\geq 2.
\end{align*}
In any case, this implies 
\begin{align*}
\lim_{m\rightarrow\infty}\tilde{P}_{N}\left(m,\left\{ m-1\right\} \right) &\geq 2\lim_{m\rightarrow\infty}\tilde{P}_{N}\left(m,\left\{ m+1\right\} \right),
\end{align*}
and since
\begin{align*}
\lim_{m\rightarrow\infty}\tilde{P}_{N}\left(m,\left\{ m-1\right\} \right) &= \min\left\{1-\theta,2\theta l\right\}\\
&> 0,
\end{align*}
the noisy chain is geometrically ergodic according to Proposition
\Cref{prop:non.hom.rw}.
\end{prooflabel}

\begin{prooflabel}[\Cref{prop:trans.geom.inhom.allN}]
Noting that
\begin{align*}
\frac{\varepsilon_{m-1}}{\varepsilon_{m}} &\in \begin{cases}
\begin{array}{c}
O\left(m^{2}\right)\\
O\left(m^{-1}\right)
\end{array}\quad\text{if} & \begin{array}{c}
m\Mod{3}=0,\\
m\Mod{3}\in\{1,2\},
\end{array}\end{cases}\\
\text{and}\quad\frac{\varepsilon_{m+1}}{\varepsilon_{m}} &\in \begin{cases}
\begin{array}{c}
O\left(m^{-2}\right)\\
O\left(m\right)
\end{array}\quad\text{if} & \begin{array}{c}
m\Mod{3}=2,\\
m\Mod{3}\in\{0,1\},
\end{array}\end{cases}
\end{align*}
expressions in \eqref{eq:ex3.expect1} and \eqref{eq:ex3.expect2}
become
\begin{align*}
\tilde{\alpha}_{N}(m,m-1) &= \left(1-s_{m-1}\right)^{N}\left(1-s_{m}\right)^{N}\mathds{1}_{\left\{ m\Mod{3}=0\right\} } +O(m^{-1})
\end{align*}
and
\begin{align*}
\tilde{\alpha}_{N}(m,m+1) &= \left(1-s_{m+1}\right)^{N}\left(1-s_{m}\right)^{N}\mathds{1}_{\left\{ m\Mod{3}=0,1\right\} } +O\left(m^{-1}\right).
\end{align*}

Therefore,
\begin{align*}
\frac{\tilde{P}_{N}\left(m,\left\{ m-1\right\} \right)}{\tilde{P}_{N}\left(m,\left\{ m+1\right\} \right)} &= \left(\frac{1-\theta}{\theta}\right)\frac{\left(1-s_{m-1}\right)^{N}+O(m^{-1})}{\left(1-s_{m+1}\right)^{N}+O(m^{-1})}\mathds{1}_{\left\{ m\Mod{3}=0\right\} }\\
 &\quad +O\left(m^{-1}\right)\mathds{1}_{\left\{ m\Mod{3}=1\right\} }\\
 &\quad +O(1)\mathds{1}_{\left\{ m\Mod{3}=2\right\} },
\end{align*}
implying there exists $C\in\mathbb{R}^{+}$ such that for $j=0,2$
\begin{align*}
\lim_{k\rightarrow\infty}\frac{\tilde{P}_{N}\left(3k+j,\left\{ 3k+j-1\right\} \right)}{\tilde{P}_{N}\left(3k+j,\left\{ 3k+j+1\right\} \right)} \leq C,
\end{align*}
and
\begin{align*}
\lim_{k\rightarrow\infty}\frac{\tilde{P}_{N}\left(3k+1,\left\{ 3k\right\} \right)}{\tilde{P}_{N}\left(3k+1,\left\{ 3k+2\right\} \right)} = 0.
\end{align*}
Then, for fixed $\delta>0$ there exists $k_{0}\in\mathbb{N}^{+}$
such that whenever $k\geq k_{0}$
\begin{align*}
\frac{\tilde{P}_{N}\left(3k+j,\left\{ 3k+j-1\right\} \right)}{\tilde{P}_{N}\left(3k+j,\left\{ 3k+j+1\right\} \right)} < C+\delta,\quad\text{for }j=0,2
\end{align*}
and
\begin{align*}
\frac{\tilde{P}_{N}\left(3k+1,\left\{ 3k\right\} \right)}{\tilde{P}_{N}\left(3k+1,\left\{ 3k+2\right\} \right)} < \delta.
\end{align*}

Let
\begin{align*}
c_{m} &:= \frac{\tilde{P}_{N}(m,\left\{ m-1\right\} )}{\tilde{P}_{N}(m,\left\{ m+1\right\} )},
\end{align*}
then for $k\geq k_{0}+1$
\begin{align*}
\prod_{j=2}^{3k+1}c_{j} &= \prod_{j=1}^{k}c_{3j-1}c_{3j}c_{3j+1}\\
&\leq \left(\left(C+\delta\right)^{2}\delta\right)^{k-k_{0}}\prod_{j=1}^{k_{0}}c_{3j-1}c_{3j}c_{3j+1}.
\end{align*}
Take $\delta$ small enough, such that $\left(C+\delta\right)^{2}\delta<1$,
hence
\begin{align*}
\sum_{k=1}^{\infty}\prod_{j=2}^{3k+1}c_{j} &= \sum_{k=1}^{k_{0}}\prod_{j=2}^{3k+1}c_{j}+\sum_{k=k_{0}}^{\infty}\prod_{j=2}^{3k+1}c_{j}\\
&\leq \sum_{k=1}^{k_{0}}\prod_{j=2}^{3k+1}c_{j}+\prod_{j=1}^{k_{0}}c_{3j-1}c_{3j}c_{3j+1} \sum_{k=k_{0}}^{\infty}\left(\left(C+\delta\right)^{2}\delta\right)^{k-k_{0}}\\
&= \sum_{k=1}^{k_{0}}\prod_{j=2}^{3k+1}c_{j}+\frac{\prod_{j=1}^{k_{0}}c_{3j-1}c_{3j}c_{3j+1}}{1-\left(C+\delta\right)^{2}\delta}\\
&< \infty.
\end{align*}
Similarly, it can be proved that
\begin{align*}
\sum_{k=0}^{\infty}\prod_{j=2}^{3k+2}c_{j} < \infty\quad\text{and}\quad\sum_{k=1}^{\infty}\prod_{j=2}^{3k}c_{j}<\infty,
\end{align*}
thus
\begin{align*}
\sum_{m=2}^{\infty}\prod_{j=2}^{m}c_{j} < \infty,
\end{align*}
implying the noisy chain is transient according to \Cref{prop:non.hom.rw}.
\end{prooflabel}

\subsection{\Cref{sec:Invariants}}
\begin{prooflabel}[\Cref{prop:rate.conv.invar}]
From \eqref{eq:aux.bound.tv}
and taking $\delta<\frac{1}{2}$, $\eta=\frac{\delta}{1-\delta}$
\begin{align*}
\sup_{x\in\mathcal{X}}\|\tilde{P}_{N}(x,\cdot) -P(x,\cdot)\|_{TV} &\leq 3\delta+4\sup_{x\in\mathcal{X}}\mathbb{P}\left[\Big\vert W_{x,N}-1\Big\vert\geq\frac{\delta}{2}\right].
\end{align*}
Using Markov's inequality
\begin{align*}
\sup_{x\in\mathcal{X}}\|\tilde{P}_{N} (x,\cdot)-P(x,\cdot)\|_{TV} &\leq 3\delta+4\sup_{x\in\mathcal{X}}\mathbb{P}\left[\Big\vert W_{x,N}-1\Big\vert^{1+k}\geq\left(\frac{\delta}{2}\right)^{1+k}\right]\\
&\leq 3\delta+\frac{2^{3+k}}{\delta^{1+k}}\sup_{x\in\mathcal{X}}\mathbb{E}\left[\Big\vert W_{x,N}-1\Big\vert^{1+k}\right]\\
&\leq 3\delta+\frac{2^{3+k}}{\delta^{1+k}N^{k}}\sup_{x\in\mathcal{X}}\mathbb{E}\left[\Big\vert W_{x}-1\Big\vert^{1+k}\right].
\end{align*}

Now, let 
\begin{align*}
C_{k} = \sup_{x\in\mathcal{X}}\mathbb{E}\left[\Big\vert W_{x}-1\Big\vert^{1+k}\right],
\end{align*}
then the convex function $f:\mathbb{R}^{+}\rightarrow\mathbb{R}^{+}$
where
\begin{align*}
f(s) = 3s+\frac{2^{3+k}C_k}{s^{1+k}N^{k}},
\end{align*}
is minimised at 
\begin{align*}
s_{*} &= \left(\frac{(1+k)2^{3+k}C_k}{3N^{k}}\right)^{\frac{1}{2+k}}\\
&= O\left(N^{-\frac{k}{2+k}}\right).
\end{align*}
Then, 
\begin{align*}
\sup_{x\in\mathcal{X}}\|\tilde{P}_{N}(x,\cdot) -P(x,\cdot)\|_{TV} &\leq f(s_{*})\\
&= O\left(N^{-\frac{k}{2+k}}\right)+O\left(N^{-k+\frac{k(1+k)}{2+k}}\right)\\
&= O\left(N^{-\frac{k}{2+k}}\right).
\end{align*}
Applying \Cref{thm:rate.conv.invar} by taking
\begin{align*}
r(N) &\propto N^{1-\frac{2}{2+k}}
\end{align*}
and noting $\log\left( N^{1-\frac{2}{2+k}} \right)\leq \log(N)$, the result is obtained.

For the second claim, for a given $\varepsilon\in(0,1)$ take $k_{\varepsilon}\geq2\left(\varepsilon^{-1}-1\right)$
and apply the first part.
\end{prooflabel}

\end{document}